\documentclass[iop]{emulateapj}
\usepackage{graphicx}
\usepackage{amssymb}
\usepackage{multirow}
\usepackage{float}
\usepackage[caption = false]{subfig}
\usepackage{comment}
\usepackage{booktabs}
\usepackage[normalem]{ulem}
\usepackage{dashrule}
\usepackage{enumitem}
\usepackage{ulem}

\newcommand{\be}{\begin{equation}}
\newcommand{\ee}{\end{equation}}

\newcommand{\chan}{{\sl Chandra }}
\newcommand{\xmm}{{\sl XMM-Newton }}

\newcommand{\grays}{$\gamma$-rays }
\def\beq{\begin{equation}}   \def\eeq{\end{equation}}
\def\beqr{\begin{eqnarray}}   \def\eeqr{\end{eqnarray}}

\begin{document}
 
\title{X-ray EMISSION FROM J1446--4701, J1311--3430 AND OTHER BLACK WIDOW PULSARS}

\author{
Prakash ~Arumugasamy,\altaffilmark{1}
George ~G.~Pavlov,\altaffilmark{1}
and Gordon ~P.~Garmire \altaffilmark{2}
}
\altaffiltext{1}{Department of Astronomy \& Astrophysics, 525 Davey Lab,
The Pennsylvania State University, University Park, PA 16802, USA}
\altaffiltext{2}{Huntingdon Institute for X-ray Astronomy, LLC, 
10677 Franks Road, Huntingdon, PA 16652, USA}
\email{pxa151@ucs.psu.edu}

\begin{abstract}
We present the results of detailed X-ray analysis of two black-widow pulsars (BWPs), J1446--4701 and J1311--3430.
PSR J1446--4701 is a BWP with orbital parameters near the median values of the sample of known BWPs.
Its X-ray emission detected by \xmm is well characterized by a soft power-law (PL) spectrum (photon index $\Gamma \approx 3$), and it shows no significant orbital modulations.
In view of a lack of radio eclipses and an optical non-detection, the system most likely has a low orbital inclination.
PSR J1311--3430 is an extreme BWP with a very compact orbit and the lowest minimum mass companion.
Our \chan data confirm the hard, $\Gamma \approx 1.3$, emission seen in previous observations.
Through phase-restricted spectral analysis, we found a hint ($\sim 2.6 \sigma$) of spectral hardening around pulsar inferior conjunction.
We also provide a uniform analysis of the 12 BWPs observed with {\sl Chandra} and compare their X-ray properties.
Pulsars with soft, $\Gamma > 2.5$, emission seem to have lower than average X-ray and $\gamma$-ray luminosities.
We do not, however, see any other prominent correlation between the pulsar's X-ray emission characteristics and any of its other properties.
The contribution of the intra-binary shock to the total X-ray emission, if any, is not discernible in this sample of pulsars with shallow observations.
\end{abstract}


\section{Introduction}

Black Widow Pulsar systems (BWPs) are compact binary systems in which a non-accreting recycled millisecond pulsar (MSP) is in the process of ablating its companion star.
The original BWP B1957+20 was discovered in the radio by observing periodic eclipses of the pulsar's emission at phases of pulsar superior conjunction, and significant dispersion measure (DM) variations immediately preceding and succeeding the eclipse \citep{1988Natur.333..237F}.
Following this discovery, \cite{1988Natur.333..832P} suggested a model for the system wherein the relativistic pulsar wind is shocked in-between the pulsar and the companion star, and the high energy radiation from the shock and the pulsar, as well as the pulsar wind, ablate the outer layers of the companion.
Observationally, the bloated and ablated outer stellar layers could produce the observed eclipse and the DM variations.
This model also predicts X-ray emission from the intra-binary shock.
Furthermore, irradiation of the companion star is expected to create temperature gradients on the stellar surface, which are observed as binary orbital modulations of the star's brightness in the optical and infrared wavelengths \citep{2007MNRAS.379.1117R}.

The model by \cite{1988Natur.333..832P} is fairly consistent with multi-wavelength observations of eclipses/modulations in the radio, X-rays and optical of some well-studied MSP binaries with short orbital periods ($< 1$ day) and very low-mass companions ($\lesssim 0.1 M_\odot$).
Hence, it has been provisionally accepted that MSPs with similar short orbits and low-mass companions must be BWPs with active stellar ablation occurring in the system.
Most of the observational differences seen in their emissions may simply be due to different viewing geometries and varying data quality.
A large fraction of the known BWP candidates are hence identified as BWPs exclusively through determination of the pulsar's orbital parameters and companion mass estimates from pulsar timing.

The periodic radio pulse delays detected near pulsar superior conjunction in very low mass MSP binaries are still the most conclusive evidence for active ablation of the companion star.
Other than B1957+20, systems that have been shown to exhibit such radio behavior are J1731--1847 \citep{2011MNRAS.416.2455B}, J2051--0827 \citep{2013MNRAS.430..571E}, J1311--3430 \citep{2013ApJ...763L..13R}, J1544+4937 \citep{2013ApJ...773L..12B} and about half a dozen globular cluster pulsars \citep{2005ASPC..328..405F}.
There have been claims for a few others (J1810+1744, J2256--1024 and J1124--3653 in \citealp{2011AIPC.1357...40H}), but the rest of the BWPs may require observations in multiple frequency bands to detect manifestations of stellar ablation.

The effects of these close encounters between MSPs and their companions are observed at multiple wavelengths, all the way up to the $\gamma$-rays.
Detections of orbital modulation in the $\gamma$-rays have been elusive for BWP systems \citep{2012ApJ...761..181W}, until very recently, when an $\sim 3\sigma$ detection of such modulation was reported for the BWP J1311--3430 \citep{2015arXiv150204783X}.
In the X-rays, Doppler boosting of intra-binary shock emission produces orbital flux modulation detected in high-inclination systems \citep{2012ApJ...760...92H}.
The pulsar spectra are expected to be similar to those of isolated MSPs: a soft power-law (photon index $\Gamma \gtrsim 2$) and/or thermal emission, with a blackbody (BB) temperature $kT \approx 0.15$ keV, presumably from the pulsar polar caps.
The relativistic intra-binary shock should, however, produce a harder, $\Gamma \approx 1.5$, power-law (PL).

In our paper, we report on latest X-ray observations of BWPs J1446--4701 and J1311--3430.
PSR J1446--4701 is a 2.2 ms recycled pulsar in a 6.7 hr orbit with a $\geq 0.019\,M_\odot$ companion (Table \ref{psrpars1}), discovered at 1.4 GHz in the radio and identified in the \grays by \cite{2012MNRAS.419.1752K} .
The pulsar, however, showed no radio eclipses in the band where it was detected.
\cite{2014MNRAS.439.1865N} further improved the timing solution and determined pulsar proper motion $\mu \approx 4.5$ mas yr$^{-1}$, using radio data over a significantly longer time baseline.

The pulsar has a spin-down power $\dot{E} = 3.6 \times 10^{34}$ erg s$^{-1}$ and an empirical intra-binary distance estimate, $\hat{a} \equiv [1+m_1/(m_2\, \sin\,i)] \times (a_1 \sin\,i) = 4.6\;{\rm lt\mbox{-} s}$, where $i$ is the orbital inclination, $m_2$ is the companion mass, $a_1$ is the semi-major axis of the pulsar orbit, and $m_1$ is the pulsar mass, assumed to be $1.4\,M_\odot$.
Since the pulsar's spin-down power and intra-binary distance were promising for detecting intra-binary shock emission in the X-rays, we conducted a 62 ks concurrent X-ray and optical observation with the \xmm EPIC and OM instruments.
We measured the X-ray flux and characterized the X-ray spectrum around superior and inferior conjunctions of the pulsar.
We also searched for an optical counterpart in the OM data.

Pulsar J1311--3430 is a member of a very short-period (94 min) binary.
Initially an unidentified {\em Fermi} source, 2FGL J1311.7--3429 was suspected to be a BWP after observations of its optical counterpart showed significant flux modulations with the binary period (\citealp{2012ApJ...757..176K}; \citealp{2012ApJ...760L..36R}).
\cite{2012Sci...338.1314P} later detected 2.56 ms $\gamma$-ray pulsations from the pulsar.
The pulsar was also detected with the Jansky Very Large Array eventually, and intermittent pulsations were found using the Green Bank Telescope \citep{2013ApJ...763L..13R}.

This system was also observed with the {\em Suzaku} and {\em Chandra} X-ray observatories.
In the 33.4 ks {\em Suzaku} observation, the pulsar showed unexpected flaring activity and some orbital modulation (\citealp{2011ApJ...729..103M}, \citealp{2012ApJ...757..176K}).
In the {\em Chandra} observation, the source's average position was in the chip gap, which significantly reduced the effective exposure \citep{2012ApJ...756...33C}.
Prior to the source identification, we had proposed a short 10 ks observation of 2FGL J1311.7--3429 with {\em Chandra} to characterize its timing and spectral properties.
Using this short exposure, in combination with the previous {\em Chandra} data, we performed basic characterization of the pulsar's X-ray emission and compared the emission in phase intervals around superior/inferior conjunctions.


\begin{deluxetable}{@{}lcc@{}}
\tabletypesize{\small}
\tablecolumns{3}
\tablewidth{0pt}
\tablecaption{Pulsar Parameters Summary (ATNF) \label{psrpars1}}
\tablehead{
\colhead{Parameter}		& \colhead{J1446--4701}	& \colhead{J1311--3430}}
\startdata
	RA (hh:mm:ss)			&14:46:35.71391(2)	&$13:11:45.7242(2)$	\\[1ex]
	DEC (dd:mm:ss)			&-47:01:26.7675(4)	&-34:30:30.350(4)	\\[1ex]
	Gl/Gb ($\degr$)			&322.500,11.425		&307.68, 28.18		\\[1ex]
	$P$ (ms)			&2.19469577985000(6)	&$2.5603710316720(3)$	\\[1ex]
	$\dot{P}$ ($10^{-20}$ s$\,{\rm s}^{-1}$)	&0.9810(2)	&2.0964(14)	\\[1ex]
	$\dot{E}$ ($10^{34}\,$erg s$^{-1}$)	&3.6		&4.9			\\[1ex]
	$\tau$ (Gyr)			&3.6			&1.94			\\[1ex]
	$B_{\rm surf}$ (MG)	&150			&234			\\[1ex]
	DM (cm$^{-3}$ pc)		&55.83202(14)		&37.84(26)		\\[1ex]
	Distance (kpc)			&1.5			&1.4			\\[1ex]
	$P_{\rm B}$ (days)			&0.27766607732(13)	&0.0651157335(7)	\\[1ex]
	$a_1 \sin i$ (lt-s)&0.0640118(3)		&0.010581(4)		\\[1ex]
	$T_{\rm asc}$ (MJD)	&55647.8044392(2)	&56009.129454(7)	\\[1ex]
	$M_{\rm C}$ ($M_\odot$)		&$\geq0.019$		&$\geq0.0082$		\\[1ex]
	Reference			&\cite{2014MNRAS.439.1865N}	&\cite{2012Sci...338.1314P}	\\[1ex]
\enddata
\tablecomments{The values of $\dot{P}$ and $\dot{P}$ dependent variables are corrected for Shklovskii effect.}
\end{deluxetable}

\section{Observation and Data Analysis}

\subsection{PSR J1446--4701}
Pulsar J1446--4701 was observed with the \xmm observatory (obsid 0693320101) on 2012 August 01 (MJD 56140) for a total of 62 ks.
EPIC-pn and MOS1/2 detectors were operated in Prime Full Window imaging mode.
The target was also observed concurrently with the Optical Monitor (OM; \citealp{2001A&A...365L..36M}).
The OM detector was operated in combined Image and Fast modes, where a central $10\farcs5 \times 10\farcs5$ window is operated in Fast mode while a larger $2\arcmin \times 2\arcmin$ window around it is operated in Image mode.
The initial 5 exposures of 4.4 ks each were taken with the V filter (510 -- 580 nm), followed by $5 \times 4.4$ ks + $5 \times 1.9$ ks exposures with the B filter (390 -- 490 nm).
The data processing was done with the \xmm Science Analysis System (SAS 13.5.0), applying standard tasks.

\begin{figure}[ht]
{\includegraphics[width=85mm,angle=0]{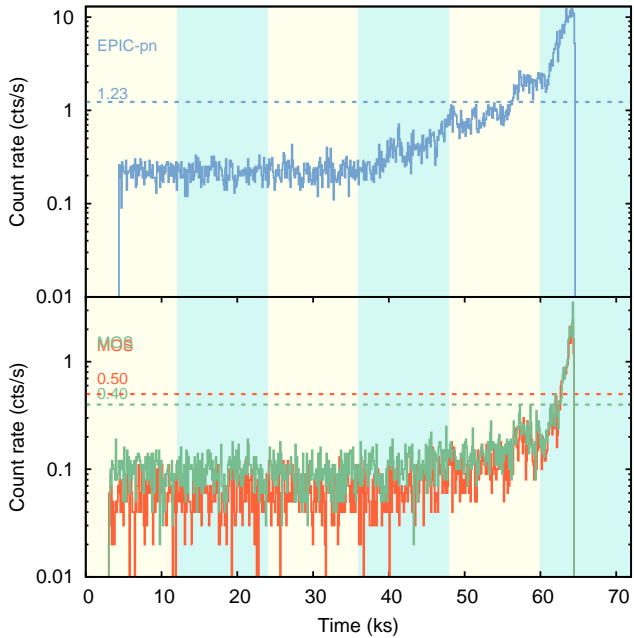}}
\caption{Flaring light curves for EPIC-pn (top), MOS1 (bottom, red) and MOS2 (bottom, green), with cut-off count rates used to obtain good time intervals (dotted).
The background colored segments separate times corresponding to half orbits centered on pulsar inferior conjunction (yellow) and superior conjunction (blue).}
\label{fig2}
\end{figure}

The EPIC observations were moderately clean, with slowly increasing soft-proton flaring background after $\approx 40$ ks of observation, and very high flaring background after 60 ks.
Periods of strong flaring are identified using light curves of single pixel events (Pattern = 0) with energies $> 10$ keV, henceforth referred to as flaring light curves\noindent\footnote{http://xmm.esac.esa.int/sas/current/documentation/threads/ EPIC\_filterbackground.shtml} (Figure \ref{fig2}).

We assigned the orbital phases to the source events using the updated orbital timing solution for the binary (Table \ref{psrpars1}; \citealp{2014MNRAS.439.1865N}).
The \xmm observation started at MJD TT 56140.83451389, 493 days ($\approx 1775$ binary orbits) after the reported epoch of ascending node, and spans over 62 ks (2.5 pulsar orbits).
The orbital period derivative is expected to be $\mathcal{O}(10^{-11}$) s s$^{-1}$ (e.g., B1957+20: \citealp{1994ApJ...426L..85A} and J2051--0827:  \citealp{2001A&A...379..579D}), and hence negligible over these time scales.

\begin{figure}[ht]
{\includegraphics[width=85mm,angle=0]{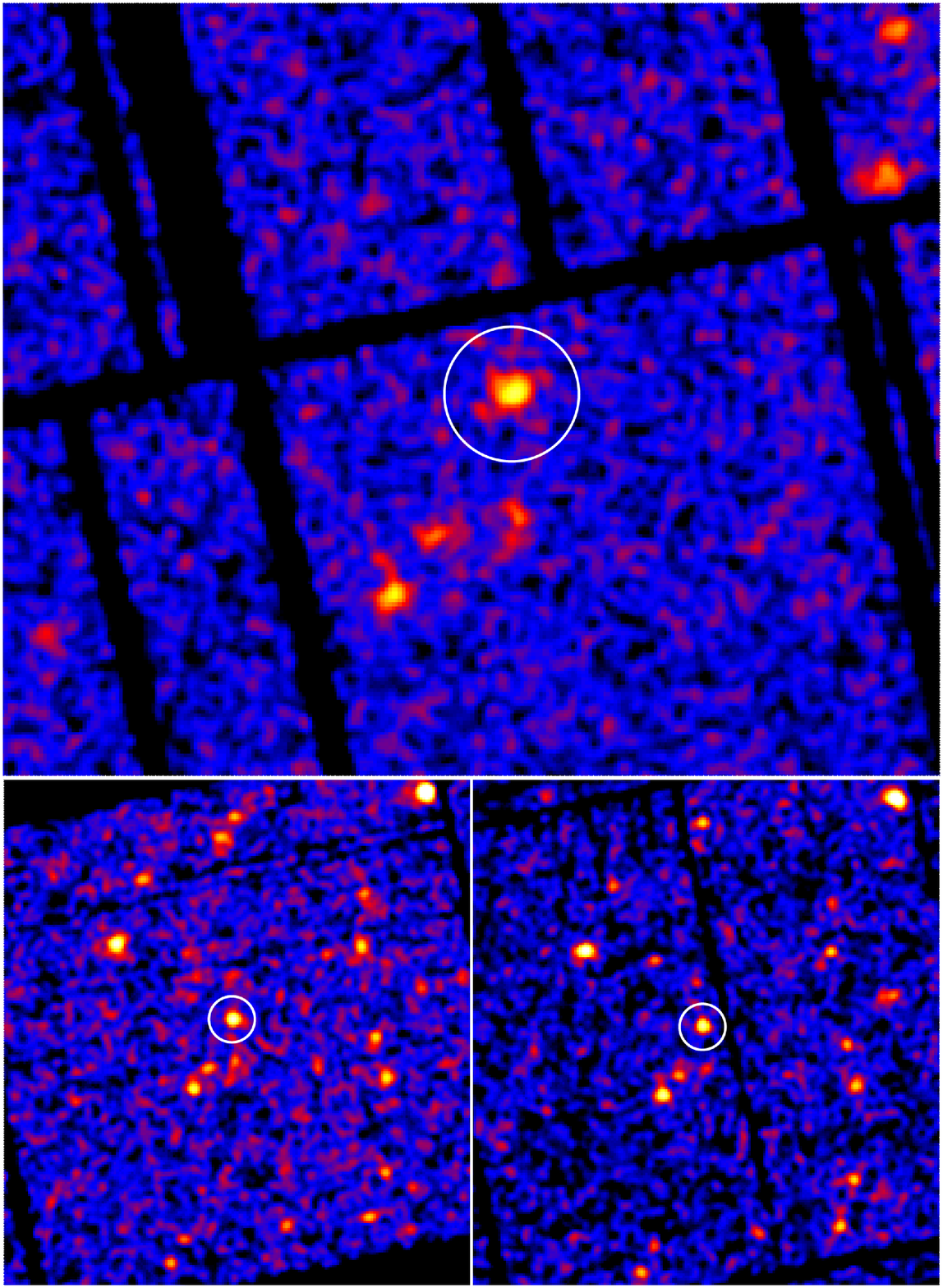}}
\caption{Top: 0.3 -- 5 keV EPIC-pn image with PSR J1446--4701 in the $30\arcsec$ radius white circle.
Bottom: 0.15 -- 5 keV MOS1 (left) and MOS2 (right) images of the same field.
North is directed up, East to the left.}
\label{fig3}
\end{figure}

Using the SAS source detection task \texttt{emldetect} on the pn image (Figure \ref{fig3}), we determined the target source coordinates, $\alpha = 14^{\rm h} 46^{\rm m} 35\fs84\pm0\fs04,\, \delta = -47^\circ 01^\prime 25\farcs5\pm0\farcs6$, with the quoted 1$\sigma$ statistical uncertainties.
This position differs from the radio position by 1\farcs8, which is well within the $4^{\prime\prime}$ average \xmm boresight error \noindent\footnote{http://xmm.esac.esa.int/external/xmm\_user\_support/ documentation/index.shtml}.

\begin{figure*}[ht]
\captionsetup[subfigure]{labelformat=empty}
\subfloat[]{\includegraphics[height=80mm,angle=0]{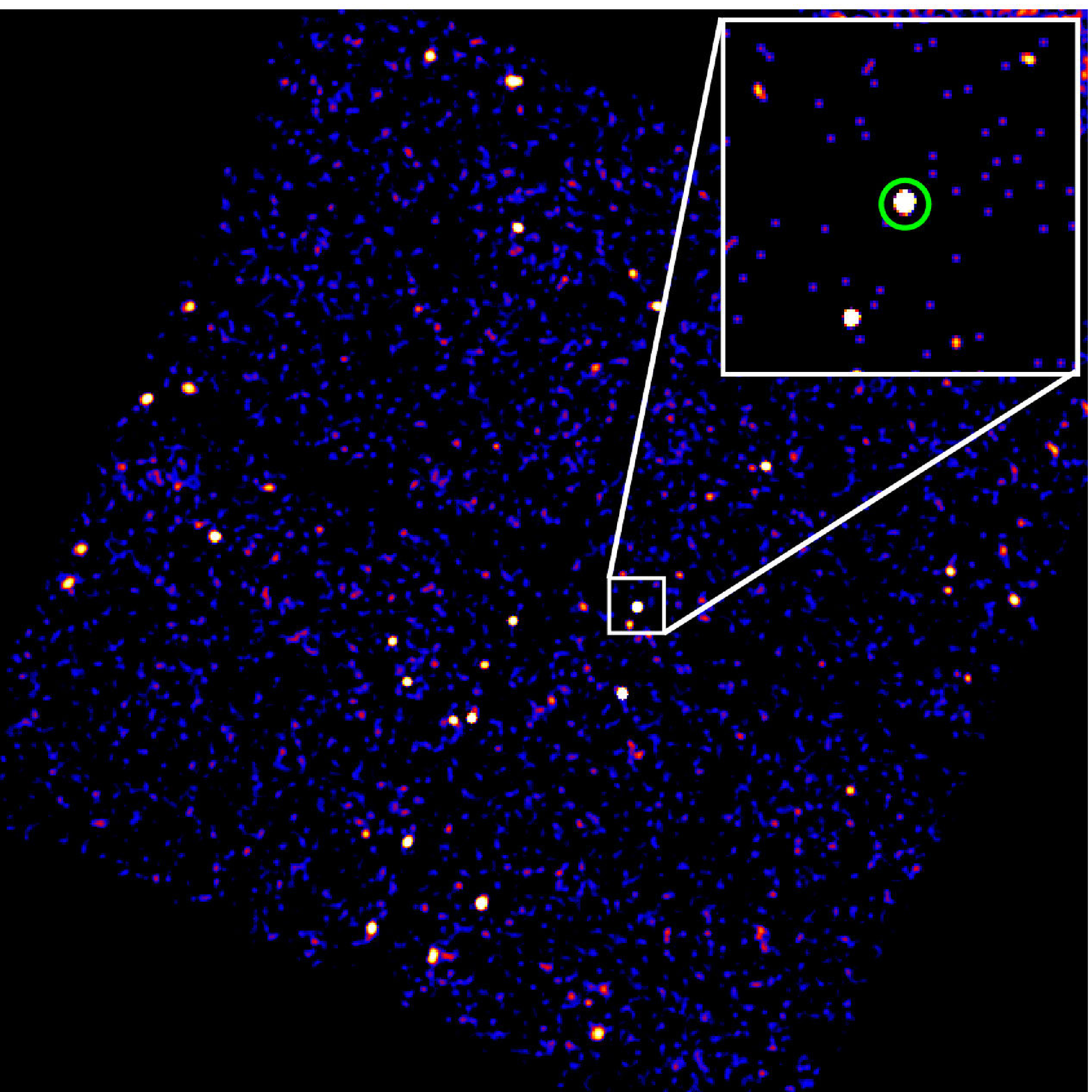}}\hspace{0.5em}
\subfloat[]{\includegraphics[height=80mm,angle=0]{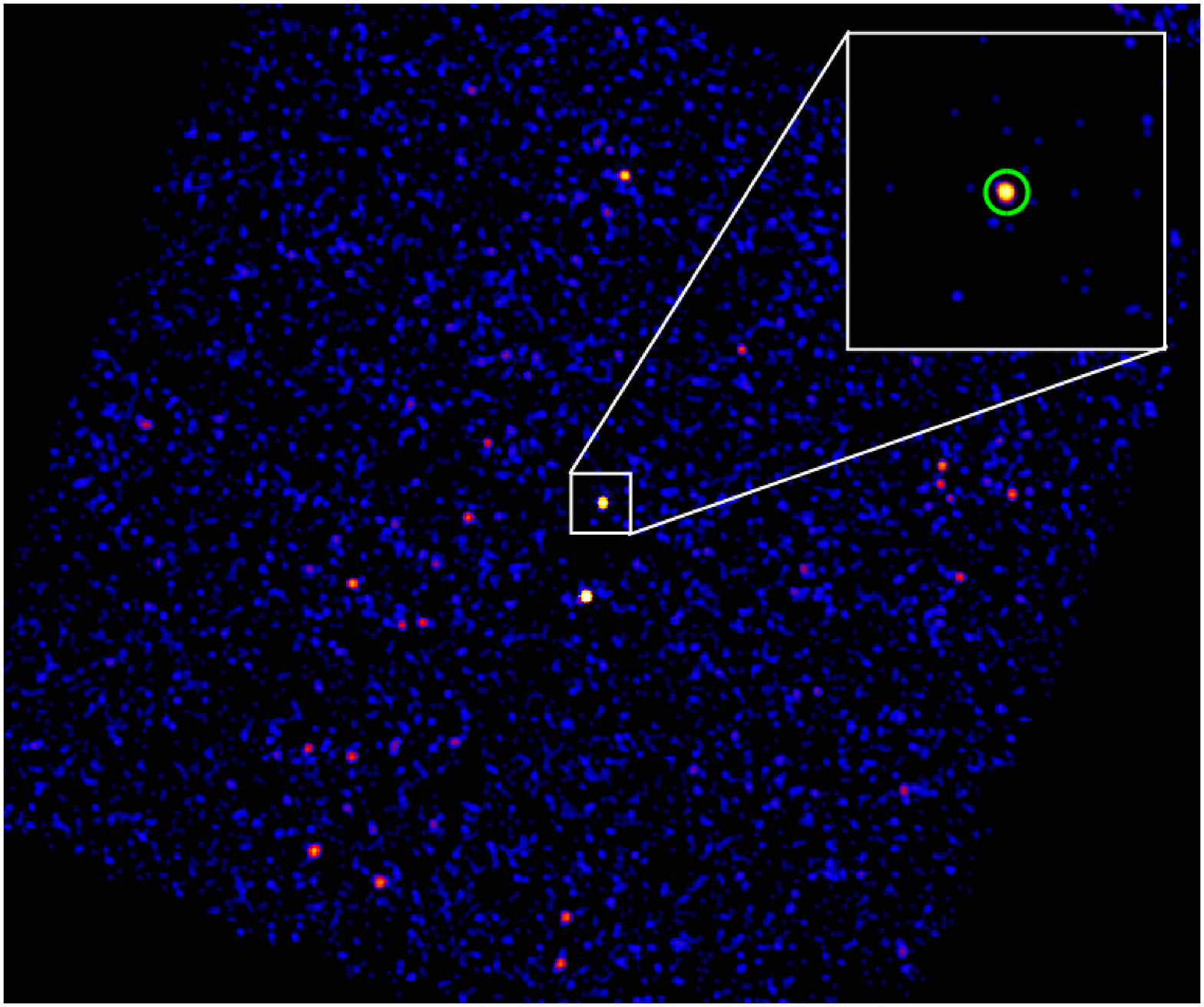}}
\caption{Smoothed 0.3 -- 7 keV ACIS-I image for ObsId 11790 (left) and ObsId 13285 (right), with PSR J1311--3430 in $1\arcmin$ box.
Insets: zoomed $1\arcmin$ regions, with $2\farcs2$ radius source extraction region shown in green.}
\label{fig1311}
\end{figure*}

\subsection{PSR J1311--3430}
We observed PSR J1311--3430 with the {\em Chandra} X-ray observatory on 2012 March 27 (MJD 56013) for 9.84 ks with the ACIS-I detector in Faint mode (ObsId 13285).
These data, along with the archival ACIS-I data (ObsId 11790, 19.96 ks) were processed with CIAO 4.7 and CALDB 4.6.5, applying the standard reprocessing script \texttt{chandra\_repro}.

To reduce background, we restricted the extracted events to the 0.3--7 keV energy range\noindent\footnote{http://cxc.harvard.edu/ciao4.4/why/filter\_energy.html}.
Using the CIAO source detection tool \texttt{wavedetect}, we detected PSR J1311--3430 at coordinates $\alpha = 13^{\rm h} 11^{\rm m} 45\fs715 \pm 0\fs004,\, \delta = -34^\circ 30^\prime 30\farcs08 \pm 0\farcs06$, which is offset from the known pulsar position (Table \ref{psrpars1}) by $0\farcs5$.
This offset is, however, within the 90\% uncertainty of {\em Chandra}'s $0\farcs6$ absolute pointing accuracy.
The source does not show any signs of extended emission, and we extracted the events from a $2\farcs2$ region around the point source, for timing and spectral analysis.

In the archival data, we chose the same energy range and extraction parameters.
Although the target position was in the chip gap for the nominal telescope pointing coordinates, $\approx 40\%$ of pulsar events were detected due to telescope dither.
The spectral information is unaffected by the telescope dither, and the reduced exposure is accounted for in the CIAO spectral extraction processes by calculating weighted responses from CCD chips on either side of the chip gap.

\section{Timing Analysis}

\subsection{PSR J1446--4701}
We performed signal-to-noise ratio (S/N) maximization over the source extraction regions and energy ranges, and extracted 313 EPIC-pn events in the 0.3 -- 5 kev range from a $13 \arcsec$ radius region.
We obtained 175 counts in the 0.15 -- 5 keV range from $13 \arcsec$ radius regions by performing similar extractions in the MOS detectors (Table \ref{specfilter}).
We used the known pulsar orbital ephemeris to assign orbital phases to the X-ray events.
Since the number of counts is low, and the data are contaminated by non-uniform background as seen from the flaring light curve, we performed unbinned analysis to look for orbital modulations.

We assume under the null hypothesis that the events from the source follow a uniform distribution over orbital phase.
The background contribution is empirically modeled using the background samples extracted from source-free regions around the pulsar position.
The source contribution is simulated by sampling from a uniform distribution.
The source plus background model distribution is then compared with the distribution of events extracted from the source region, using the Anderson-Darling test  (AD; \citealt{doi:10.1080/01621459.1974.10480196}).

\begin{figure*}[ht]
\captionsetup[subfigure]{labelformat=empty}
\subfloat[]{\includegraphics[width = 90mm]{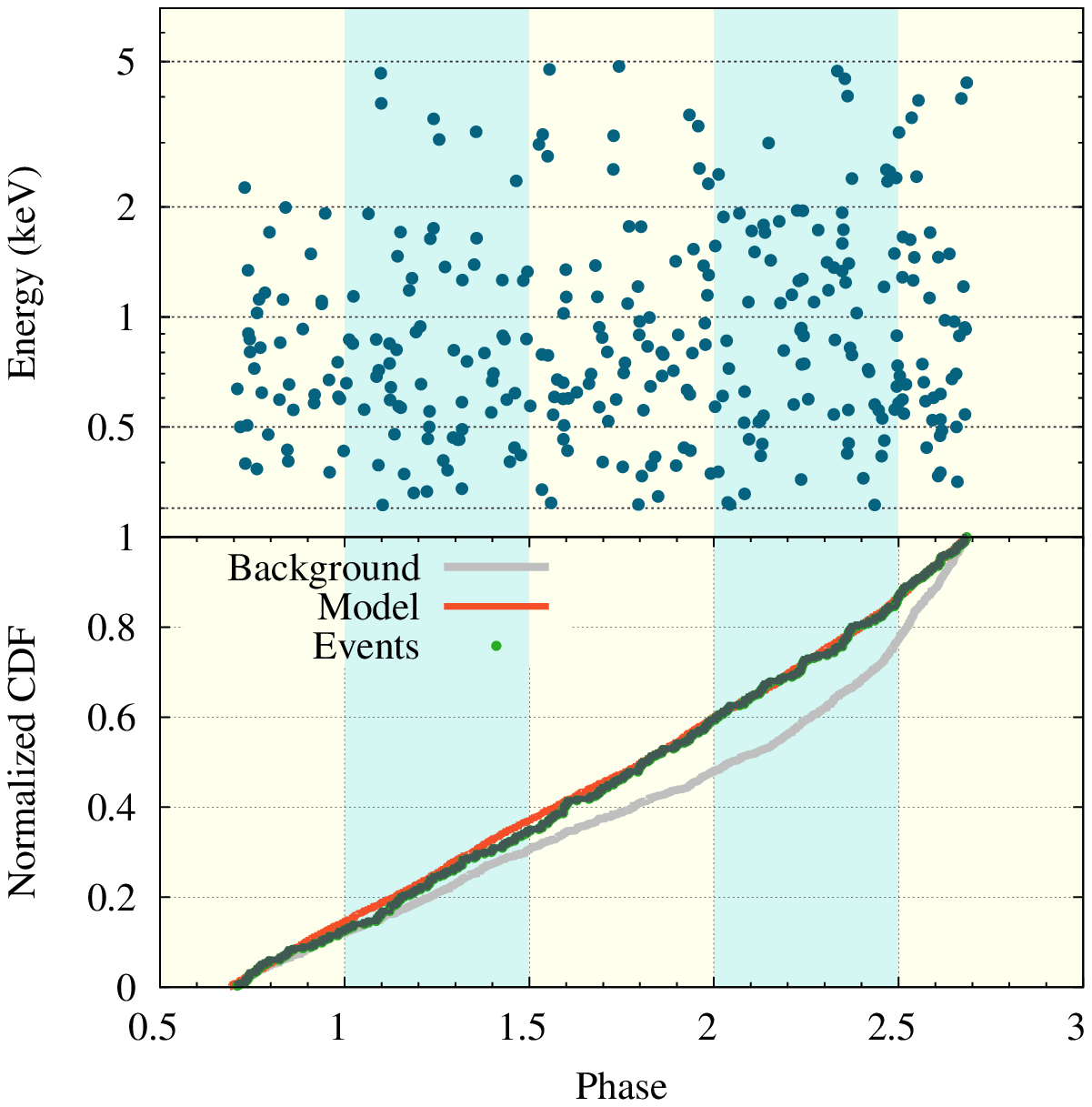}}
\subfloat[]{\includegraphics[width = 90mm]{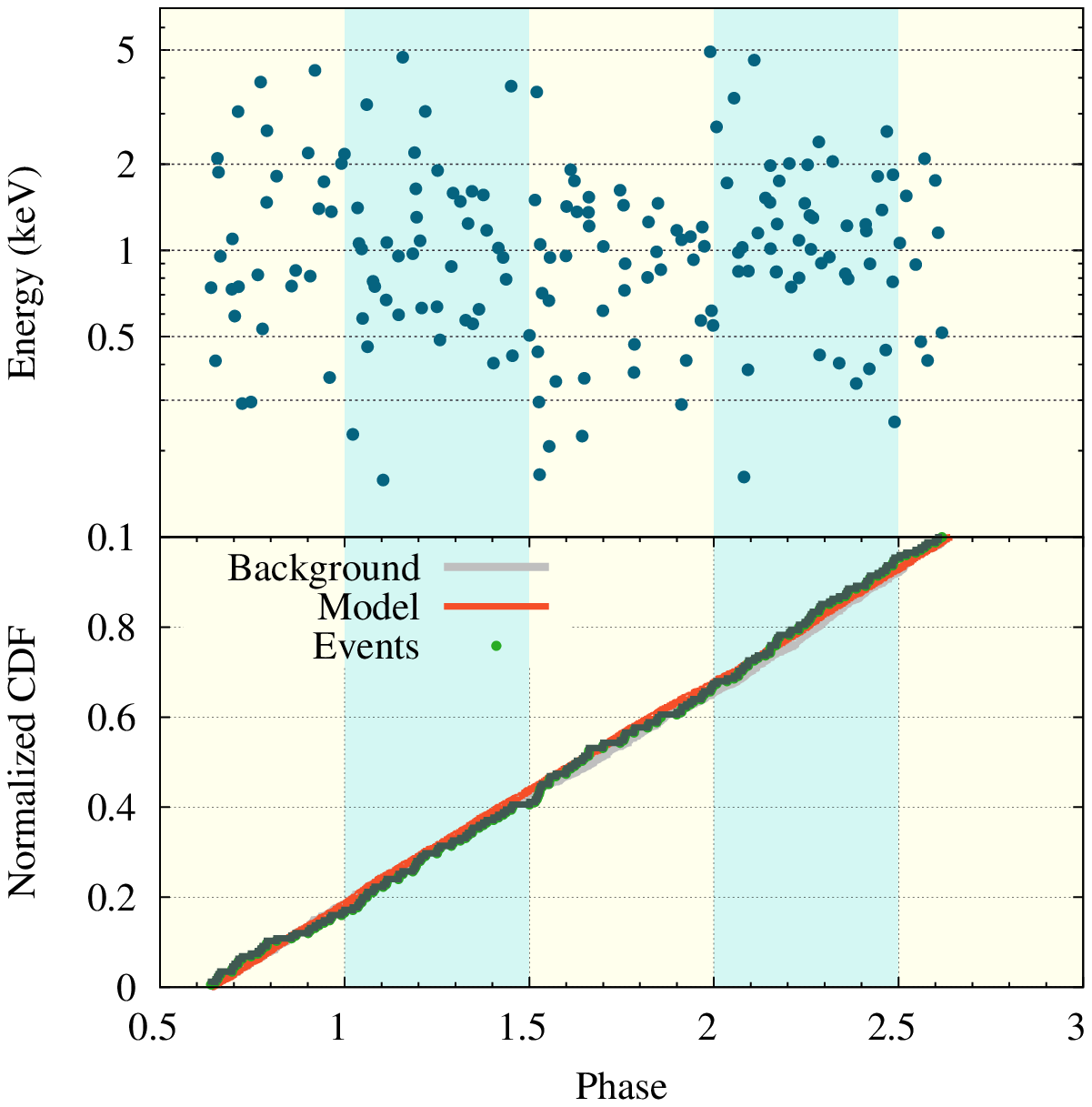}}
\caption{X-ray photon energy versus orbital phase ($\phi=0.0$ corresponds to ascending node) over 2 complete orbits of the BWP J1446--4701 for EPIC-pn ($N_{\rm 0.3-5\,keV} = 313$ counts), shown left-top, and combined MOS1/2 ($N_{\rm 0.15-5\,keV} = 175$ counts), shown right-top.
Background shades cover half orbits centered at the pulsar inferior conjunction (yellow) and superior conjunction (blue).
Normalized CDF of source events' distribution (green) compared with the background plus uniform source distribution model (red), as a precursor to the AD test (bottom panels). The normalized CDF of extracted background events are also plotted for reference (grey).}
\label{fig4}
\end{figure*}

We use R \citep{Rprogram} implementation of the AD test in the \texttt{kSamples} package \citep{kSamplesPackage}.
At p-values of 0.86 (pn) and 0.9 (MOS), the null hypotheses that the photon distributions follow an underlying uniform distribution are not rejected.
The normalized cumulative distribution functions (CDFs) for the background, simulated uniform source plus background, and the extracted source region events, for pn and MOS are plotted in Figure \ref{fig4}.
The source events do not show statistically significant deviations from a uniform distribution.

\subsection{PSR J1311--3430}
\begin{figure}[H]
\includegraphics[width = 85mm]{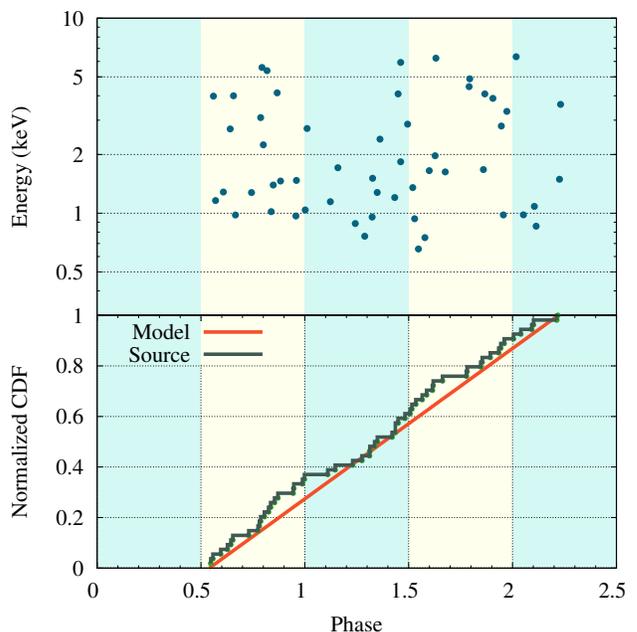}
\caption{PSR J1311--3430: X-ray photon energy versus orbital phase ($\phi=0.0$ corresponds to ascending node) over $\approx 1.67$ pulsar orbits ($N_{\rm 0.3-7\,keV} = 54$ counts) in top panel.
Background shades cover half orbits centered at the pulsar inferior conjunction (yellow) and superior conjunction (blue).
Normalized CDF of source events' distribution compared with a uniform distribution as a precursor to the AD test (bottom panel).}
\label{fig5}
\end{figure}

For ObsID 13285, we extracted 54 counts (background contribution is $< 1$ count) from a $2\farcs2$ radius circle around the source position.
The 9.8 ks exposure corresponds to 1.67 orbits of the binary.
Due to the low number of counts, binning or smoothing to look for timing variations is not feasible.
In Figure \ref{fig5}, we show phases versus energy for 0.3--7 keV events.
There is a hint of hard ($> 2$ keV) X-ray photons depletion near superior conjunction of the pulsar (phase ranges: 1--1.5, 2--2.25), or conversely, an excess of hard X-rays around inferior conjunction (phase ranges: 0.5--1, 1.5--2).
The difference in count rates at different orbital phases cannot be well constrained due to relatively large uncertainties.

We model the source events with a uniform distribution and perform the AD test to test the null hypothesis.
The null hypothesis is not rejected (p-value = 0.62).
Although, there are hints of local deviations between the source events and the uniform distribution, they seem to be of low statistical significance (Figure \ref{fig5}).

\section{Spectral Analysis}

We analyze phase-integrated spectra covering the entire orbital phase range.
We also analyze the spectra restricted to two broad phase ranges, half orbit around pulsar superior conjunction ($\phi= 0.0 - 0.5$) and inferior conjunction ($\phi= 0.5 - 1.0$), henceforth referred to as SC phases (or SCP) and IC phases (or ICP), respectively.

\subsection{PSR J1446--4701}
Although the pulsar's 0.3 -- 5 keV X-ray emission exhibits no significant count-rate modulations, one cannot exclude a priori the possibility that the spectrum is phase-dependent.
To check for spectral variations, we compared the spectral fits for IC and SC phases.
We maximize S/N of the spectra by choosing optimal extraction aperture sizes and exposure intervals.
For phase-integrated spectra, we maintain equal contributions from the IC and SC phases, by restricting the extraction interval to an integral number of orbital periods (2 orbits).
The spectral extraction parameters are listed in Table \ref{specfilter}.

\begin{deluxetable*}{@{}lccccccccc@{}}[p]
\tabletypesize{\footnotesize}
\tablecolumns{10}
\tablewidth{0pt}
\tablecaption{Extraction parameters for orbital phase-integrated and phase-resolved spectra for PSR J1446--4701.\label{specfilter}}
\tablehead{
\colhead{}	& \colhead{}	&\colhead{Full (0--1)}	& \colhead{}	& \colhead{}	&\colhead{IC (0.5--1)}	& \colhead{}	&\colhead{}	& \colhead{SC (0--0.5)}	&\colhead{}	\\
\colhead{} & \colhead{EPIC-pn}	& \colhead{MOS1}		& \colhead{MOS2}		& \colhead{EPIC-pn}	& \colhead{MOS1}		& \colhead{MOS2}		& \colhead{EPIC-pn}	& \colhead{MOS1}		& \colhead{MOS2} }
\startdata
Energy range (keV)	  & 0.3 -- 5 	& 0.15 -- 5		& 0.15 -- 5		& 0.3 -- 5 	& 0.15 -- 5		& 0.15 -- 5		& 0.3 -- 5 	& 0.15 -- 5		& 0.15 -- 5	\\[1.5ex]
Exposure (ks)	  & 42.64	& 47.38		& 47.41		& 24.77		& 32.44		& 32.48	& 21.34		& 26.21		& 26.21		\\[1.5ex]
Extraction radius & $13\arcsec$	& $13\arcsec$	& $13\arcsec$	& $13\arcsec$	& $13\arcsec$	& $13\arcsec$	& $15\arcsec$	& $12\arcsec$	& $14\arcsec$	\\[1.5ex]
Total Counts	  & 313		& 85		& 90	  	& 203		& 69		& 50	& 177		& 48		& 59		\\[1.5ex]
Count rate (cts/ks) & 5.55 $\pm$ 0.43& 1.40 $\pm$ 0.20	& 1.52 $\pm$ 0.20 & $5.46 \pm 0.60$& $1.64 \pm 0.26$ & $1.10 \pm 0.22$	& $6.05 \pm 0.65$	& $1.32 \pm 0.26$ & $1.70 \pm 0.30$\\[1.5ex]
\enddata
\end{deluxetable*}

We used XSPEC 12.8.2\footnote{\url{http://heasarc.gsfc.nasa.gov/docs/xanadu/xspec}} for X-ray spectral analysis.
We modeled absorption by the interstellar medium (ISM) using the T\"{u}bingen-Boulder model \citep{2000ApJ...542..914W} through its XSPEC implementation \texttt{tbabs}, setting the abundance table to \texttt{wilm} \citep{2000ApJ...542..914W} and photoelectric cross-section table to \texttt{bcmc} \citep{1992ApJ...400..699B}, with updated He photo-ionization cross-section based on \citet{1998ApJ...496.1044Y}.

\begin{figure*}[ht]
\captionsetup[subfigure]{labelformat=empty}
\subfloat[]{\includegraphics[width = 85mm]{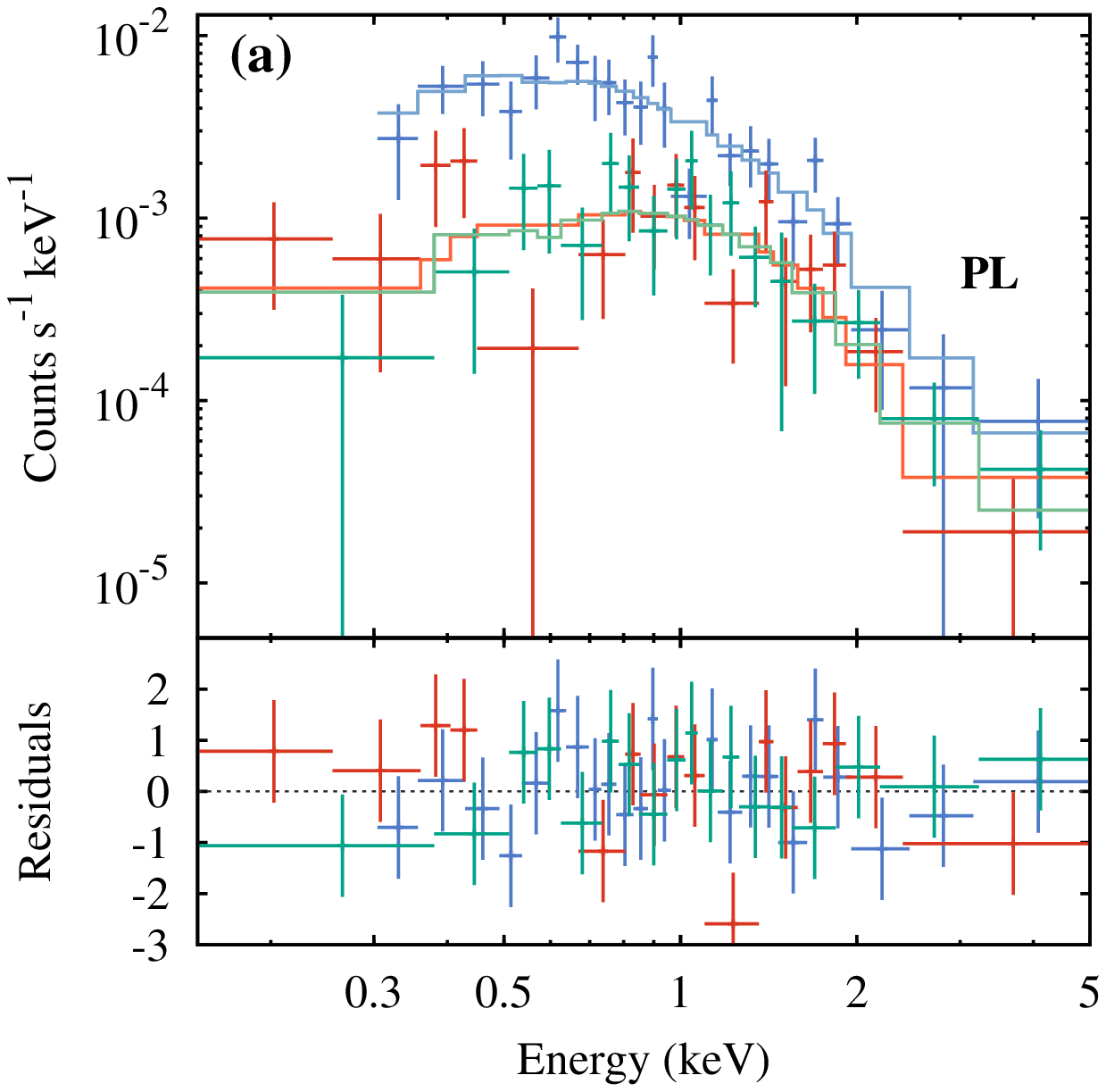}}
\subfloat[]{\includegraphics[width = 85mm]{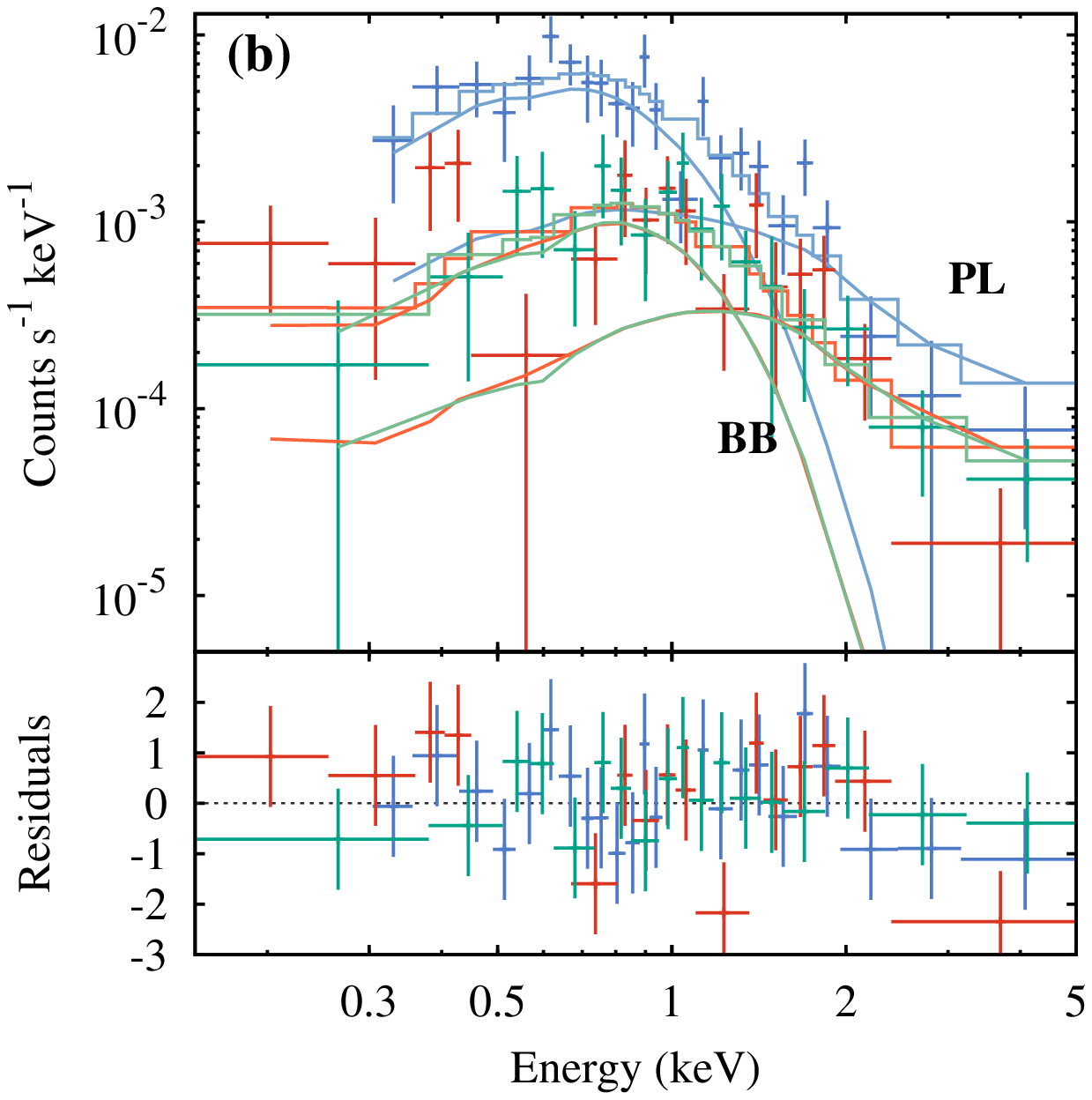}} \\[-5EX]
\subfloat[]{\includegraphics[width = 85mm]{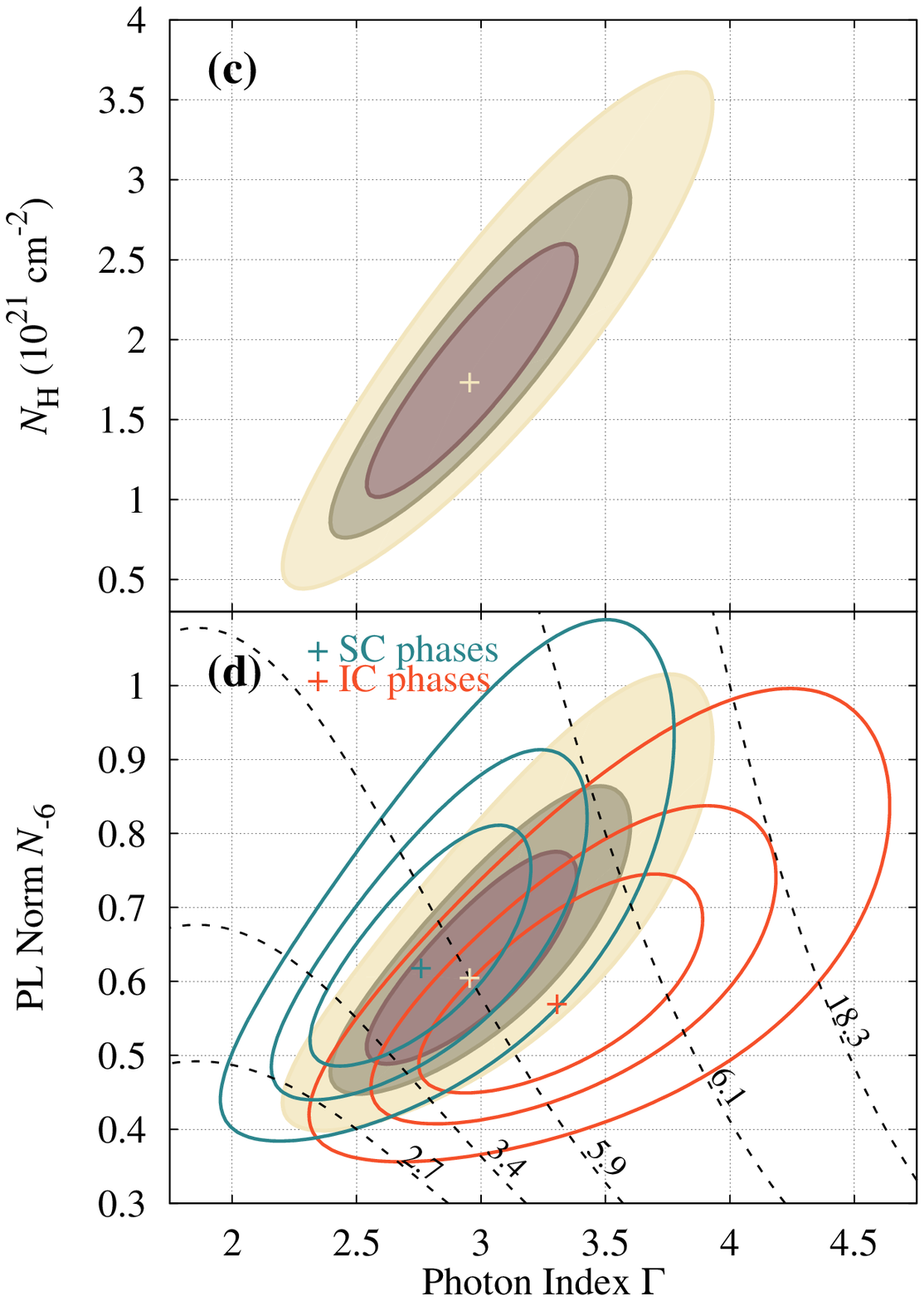}}
\subfloat[]{\includegraphics[width = 85mm]{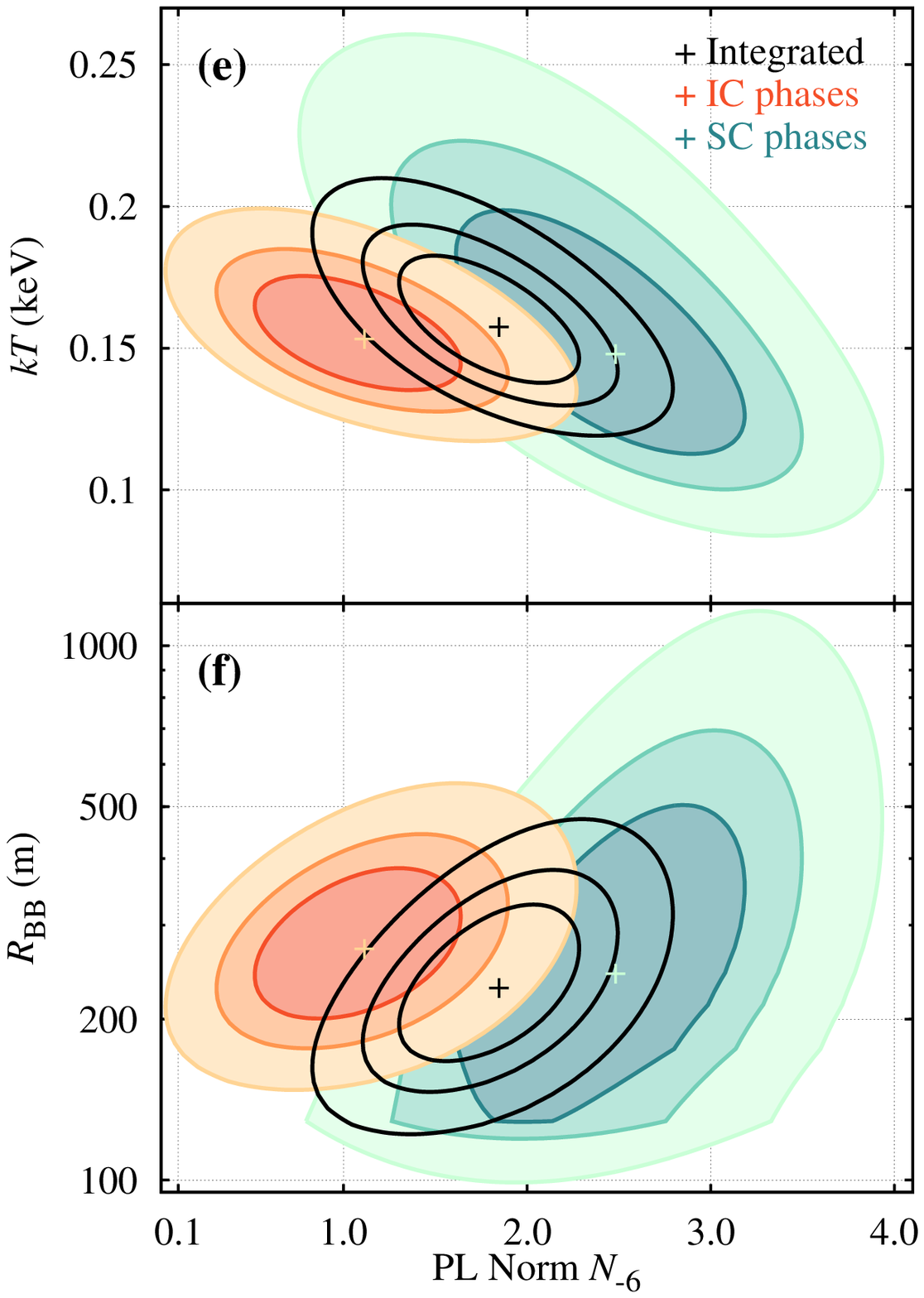}}
\caption{
Spectral analysis of PSR J1446--4701.
Panels (a) and (b) show PL and PL+BB fits, respectively, for 0.15--5 keV MOS (MOS1 in red and MOS2 in green) and 0.3 -- 5 keV pn (blue) phase-integrated spectra.
Individual component contributions of PL+BB fits are shown with continuous lines.
The fit residuals are defined as $(D-M)/\sigma_D$, where $D$ is the data counts, $M$ is the model counts in a particular bin, and $\sigma_D$ is the uncertainty in the data.
Confidence contours (68\%, 90\% and 99\%) of parameters $\Gamma - N_{\rm H}$ and $\Gamma - N_{-6}$ are shown in panels (c) and (d), respectively, for PL fit to the phase-integrated spectra of J1446--4701.
Corresponding $\Gamma - N_{-6}$ contours for SC (blue) and IC (red) phases' spectral fits are overplotted in panel (d), along with lines of constant PL flux in units of $10^{-14}$ erg cm$^{-2}$ s$^{-1}$ (black).
Panels (e) and (F) show the confidence contours of parameters $N_{-6} - kT$ and $N_{-6} - R_{\rm BB}$, respectively, for combined PL + BB fits to SC (blue-green), IC (red-orange), and full (black) phases spectra.
}
\label{J1446contours}
\end{figure*}

We binned the pn and MOS spectra with a minimum of 10 and 5 counts per bin, respectively, and adopted C-statistics \citep{1979ApJ...228..939C} for parameter estimation.
PL models with photon index $\Gamma \approx 3$ provide good fits to SC, IC and phase-integrated spectra (Table \ref{specfitPLBB}, Figure \ref{J1446contours}).
SC spectra are slightly harder, and show a higher flux compared to IC spectra, but the differences are still within the 90\% statistical uncertainty limits.

\begin{deluxetable}{@{}lccc@{}}[h]
\tabletypesize{\footnotesize}
\tablecolumns{4}
\tablewidth{0pt}
\tablecaption{Best fit spectral parameters with 90\% confidence uncertainties for PSR J1446-4701.\label{specfitPLBB}}
\tablehead{
\colhead{Parameter}	&	\colhead{Full orbit}	&	\colhead{IC (0.5--1)}	& \colhead{SC (0--0.5)}}
\startdata
\multicolumn{4}{l}{PL fit} \\
\cmidrule(r){1-1} \\[-1ex]
$N_{\rm H}$ ($10^{22}$ cm$^{-2}$)	& $0.17^{+0.10}_{-0.08}$	& \multicolumn{2}{c}{$0.17^{+0.10}_{-0.08}$}	\\[1.5ex]
$\Gamma$				& $2.93^{+0.50}_{-0.42}$		& $3.27^{+0.68}_{-0.56}$		& $2.72^{+0.52}_{-0.44}$	\\[1.5ex]
PL norm ($N_{-6}$)\tablenotemark{a}	& $6.1^{+1.8}_{-1.3}$	& $5.7^{+1.9}_{-1.3}$		& $6.2^{+2.1}_{-1.5}$	\\[1.5ex]
Cstat/d.o.f.				& $55.8/56$	&	 \multicolumn{2}{c}{50.48/73}	\\[1.5ex]
$\chi_\nu^2$				& $1.13$	&	 \multicolumn{2}{c}{0.64}	\\[1.5ex]
$F^{\rm abs}_{0.15-5\,{\rm keV}}$\tablenotemark{b}	&	$1.36^{+0.18}_{-0.16}$	&	$1.25^{+0.22}_{-0.19}$		& $1.45^{+0.25}_{-0.22}$	\\[1.5ex]
$F^{\rm unabs}_{0.15-5\,{\rm keV}}$\tablenotemark{b}	&	$5.9^{+6.9}_{-2.5}$	&	$7.9^{+15.5}_{-4.2}$		& $5.0^{+5.6}_{-2.0}$	\\[1.5ex]
\midrule
\multicolumn{4}{l}{PL+BB fit} \\
\cmidrule(r){1-1} \\[-1ex]
$N_{\rm H}$ ($10^{22}$ cm$^{-2}$)	& $0.15^{\rm F}$	&	$0.15^{\rm F}$		& $0.15^{\rm F}$	\\[1.5ex]
$\Gamma$				& $1.5^{\rm F}$	&	$1.5^{\rm F}$		& $1.5^{\rm F}$		\\[1.5ex]
PL norm ($N_{-6}$)\tablenotemark{a}	& $1.79^{+0.54}_{-0.52}$	&	$1.05^{+0.64}_{-0.58}$		& $2.42^{+0.82}_{-0.88}$	\\[1.5ex]
$F^{\rm unabs, PL}_{0.15 - 5\,{\rm keV}}$\tablenotemark{b}	& $1.06^{+0.32}_{-0.31}$	&	$0.62^{+0.38}_{-0.34}$	&	$1.44^{+0.49}_{-0.52}$	\\[1.5ex]
$kT$ (keV)				& $0.16^{+0.03}_{-0.02}$	&	$0.15^{+0.02}_{-0.02}$		& $0.15^{+0.05}_{-0.04}$	\\[1.5ex]
BB Norm \tablenotemark{c}		& $2.4^{+2.7}_{-1.2}$	&	$3.7^{+3.6}_{-1.7}$		& $2.7^{+10.1}_{-2.1}$	\\[1.5ex]
$R_{\rm BB}$ (m)\tablenotemark{c}	& $232^{+105}_{-70}$	&	$275^{+120}_{-80}$		& $247^{+289}_{-126}$	\\[1.5ex]
$F^{\rm unabs, BB}_{0.15 - 5\,{\rm keV}}$\tablenotemark{b}	&	$1.59^{+0.33}_{-0.28}$	&	$1.99^{+0.45}_{-0.38}$		& $1.39^{+0.63}_{-0.42}$	\\[1.5ex]
Cstat/d.o.f.				& $64.8/56$	&	$22.0/38$		& $27.70/33$	\\[1.5ex]
$\chi_\nu^2$				& $1.28$	&	$0.56$		& $0.81$	\\[1.5ex]
$F^{\rm abs}_{0.15 - 5\,{\rm keV}}$\tablenotemark{b}		&	$1.45^{+0.19}_{-0.18}$	&	$1.26^{+0.23}_{-0.21}$		& $1.60^{+0.29}_{-0.28}$	\\[1.5ex]
$F^{\rm unabs}_{0.15 - 5\,{\rm keV}}$\tablenotemark{b}	&	$2.65^{+0.45}_{-0.39}$	&	$2.61^{+0.57}_{-0.47}$		& $2.82^{+0.87}_{-0.64}$	\\[1.5ex]
\enddata
\tablecomments{IC/SC spectra are simultaneously fit with PL, with their $N_{\rm H}$ tied-up.}
\tablenotetext{a}{PL normalization in units of $10^{-6}$  photons cm$^{-2}$ s$^{-1}$ keV$^{-1}$ at 1 keV.}
\tablenotetext{b}{$F^{\rm abs}_{0.15-5\,{\rm keV}}$ and $F^{\rm unabs}_{0.15-5\,{\rm keV}}$ are absorbed and unabsorbed fluxes, respectively, in units of $10^{-14}$ erg cm$^{-2}$ s$^{-1}$.}
\tablenotetext{c}{${\rm BB\,Norm} = R_{\rm BB}^2/d_{10\, {\rm kpc}}^2$, where $d_{10\,{\rm kpc}}$ is the distance in units of 10 kpc, and $R_{\rm BB}$ is the BB emission effective radius in km. $R_{\rm BB}$ values are calculated assuming $d_{10\,{\rm kpc}} = 0.15$.}
\tablenotetext{F}{Parameter fixed.}
\end{deluxetable}

We also fit combined BB and PL models to the SC, IC and phase-integrated spectra.
The multi-parameter model does not sufficiently constrain the fit parameters, so first we fix the hydrogen column density at $N_{\rm H} = 1.5 \times 10^{21} {\rm cm}^{-2}$, obtained from the known dispersion measure of the pulsar (DM = 55.8 cm$^{-3}\,$pc), assuming $\approx 10\%$ ionization \citep{2013ApJ...768...64H}.
Then we fix the photon index at $\Gamma = 1.5$, a value expected from non-thermal emission from intra-binary shocks, and model the thermal emission.
In each case, the best-fit BB temperatures are $kT \approx 0.15$ keV, and effective BB radii $R_{\rm BB} \sim 250$ m (Table \ref{specfitPLBB}, Figure \ref{J1446contours}).

The BB temperatures and fluxes are fairly similar at the two opposing phase ranges.
The non-thermal emission parameters, however, show a higher degree of variation (Figure \ref{J1446contours}).
We observe lower normalization and flux value for ICP than for SCP but, due to rather large statistical uncertainties, the two normalization values are just within $2.1\,\sigma$ of each other.

\subsection{PSR J1311--3430}
Despite the small number of counts ($\approx 50$) in each observation (ObsId 11790 and 13285), a crude X-ray spectral analysis for J1311--3430 is possible due to negligible background contributions.
For ObsId 11790, the effective exposure of 8.27 ks is a fraction of the total 19.86 ks exposure, calculated from the reduced effective area for the source position (Table \ref{J1311specfilter}).

Due to the small number of counts, we use the simple PL model to characterize the spectral slope and look for spectral variations between the two observation epochs (MJD 55276 and MJD 56013) and the two orbital phase ranges.
The spectra from the two epochs have similar spectral indices and PL normalization, hence we tied-up their model parameters and fit their spectra simultaneously (Table \ref{specfitsJ1311}).
This helps improve the fit and lower the parameter uncertainties.

A $\Gamma = 1.3\pm0.3$ model provides a good fit to the phase-integrated spectra with the free absorption hydrogen column density parameter.
The $N_{\rm H} <  1.4 \times 10^{21}\;{\rm cm}^{-2}$, as constrained by the fit, is consistent with $N_{\rm H} = 1.2 \times 10^{21}\,{\rm cm}^{-2}$ estimated from the pulsar's dispersion measure (DM = 37.8 cm$^{-3}\,$pc), assuming $\approx 10\%$ ionization.
However, the estimated total Galactic neutral hydrogen column density along the pulsar line of sight is just $\approx 5 \times 10^{20}\,{\rm cm}^{-2}$ \citep{2005A&A...440..775K}.
We chose to fix $N_{\rm H} = 5 \times 10^{20}\,{\rm cm}^{-2}$ while fitting the phase-integrated and phase-resolved spectral models to further improve constraints on the PL fit parameters.

The SCP and ICP count rates are listed in Table \ref{J1311specfilter}, and their PL fit parameters in Table \ref{specfitsJ1311}.
We first compared the spectra obtained from IC phases and SC phases non-parametrically (Figure \ref{J1311spectrals}b), using a two-sample KS-test \citep{BIMJ:BIMJ19730150311} implemented in the statistical package R \citep{rdct:r}.
The hypothesis that the SCP and ICP spectra are extracted from the same underlying distribution in both epochs, is not rejected with high significance (p-value = 0.09).
The spectral fitting shows that ICP spectra ($\Gamma \approx 1$) are harder than the SCP spectra ($\Gamma \approx 2$) at $2.6\sigma$ confidence.

\section{Optical limits for J1446--4701}
We did not find any source at the target location, in neither the B- nor V-filter, for stacked or unstacked exposures obtained with \xmm OM.
For a $5\sigma$ detection in the stacked 22 ks V-filter exposure, the limiting magnitude estimate is $m_{\rm V} = 20.1$, and the faintest $3\sigma$ detection in the exposure has an estimated $m_{\rm V} = 22.1\pm0.4$.
A similar $5\sigma$ detection limit in the stacked 31.29 ks B-filter exposure is $m_{\rm B} = 21.4$, and the faintest $3\sigma$ detection has a magnitude of $m_{\rm B} = 23.5\pm0.4$.
We adopt $3\sigma$ magnitude lower limits $m_{\rm B} > 23.5$ and $m_{\rm V} > 22.1$ for the companion star to PSR J1446--4701.

\begin{figure*}[ht]
\captionsetup[subfigure]{labelformat=empty}
\subfloat[]{\includegraphics[width = 90mm]{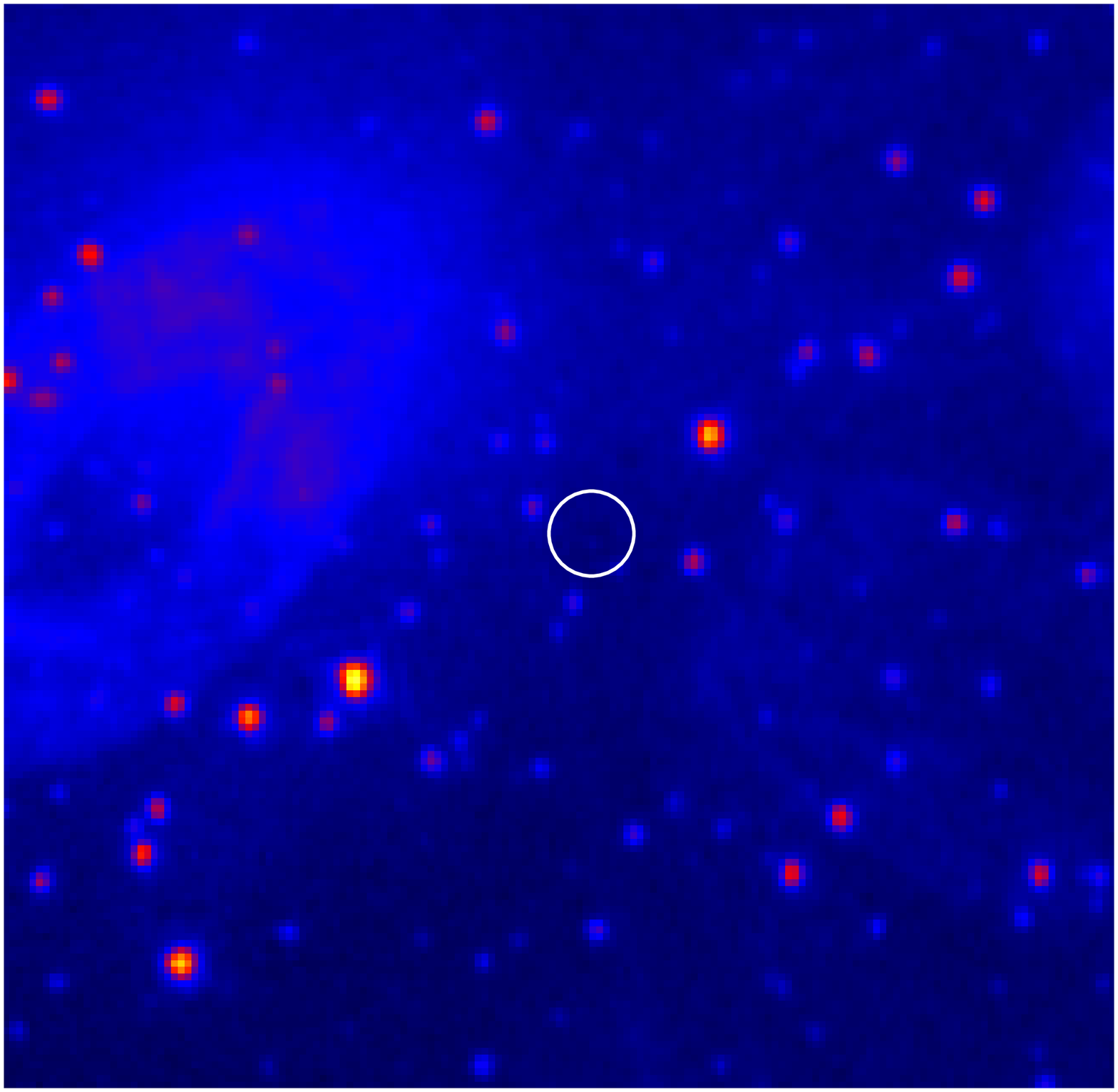}}
\subfloat[]{\includegraphics[width = 90mm]{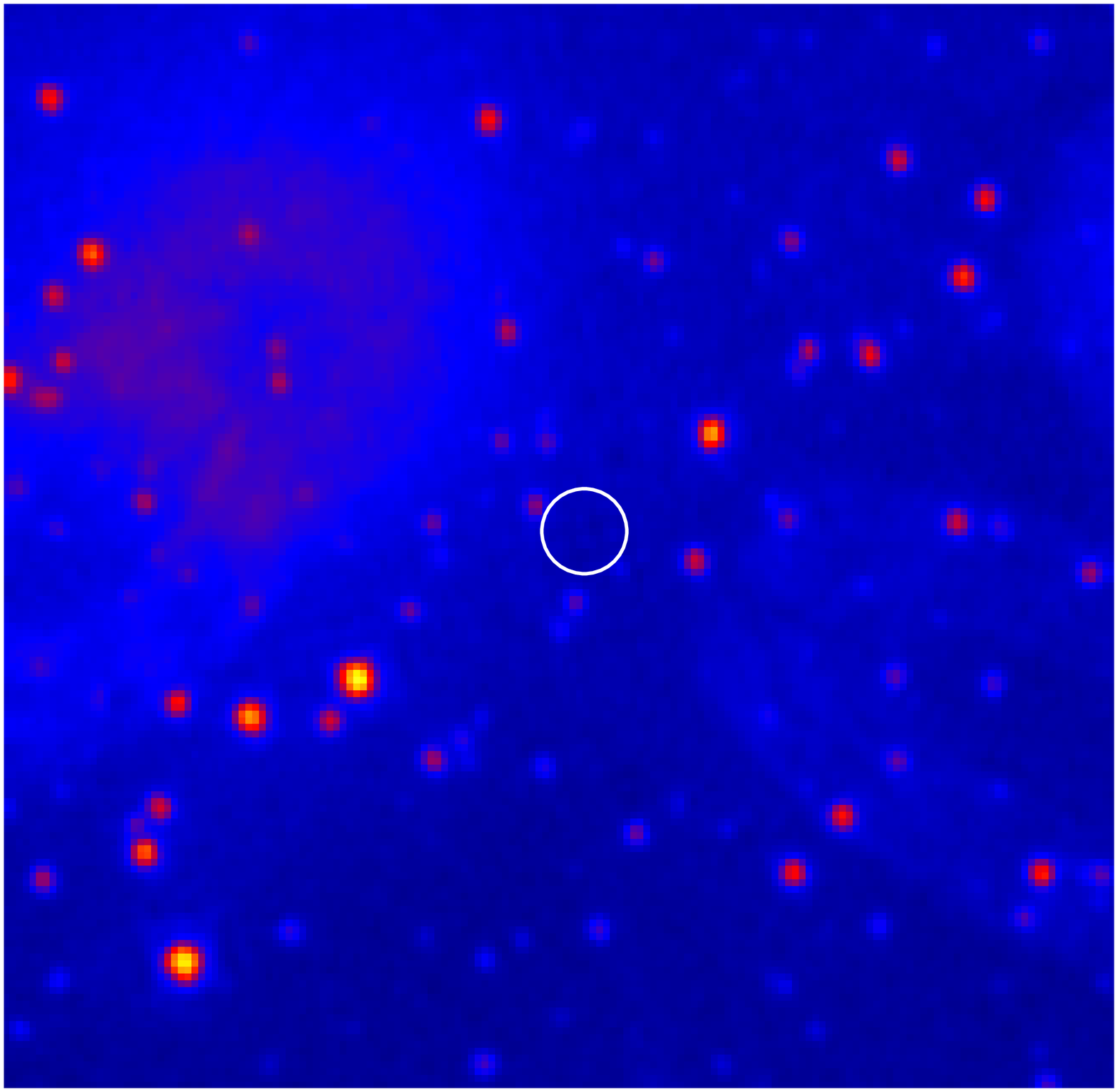}} \\
\caption{Stacked, $5 \times 4.4$ ks V-band image (left) and $5 \times 4.4$ ks + $5 \times 1.9$ ks B-band image (right).
The $6\arcsec$ radius white circles show the location of BWP J1446--4701 in the $2\farcm5\times2\farcm5$ region with north oriented towards the top.
}
\label{J1446OM}
\end{figure*}

\begin{deluxetable*}{@{}lcccccc@{}}[p]
\tabletypesize{\footnotesize}
\tablecolumns{7}
\tablewidth{0pt}
\tablecaption{Count rates for J1311--3430 orbital phase-integrated and phase-resolved data.\label{J1311specfilter}}
\tablehead{
\colhead{Phase range}	&	\colhead{Full (0--1)}\hspace{-2cm} & \colhead{} & \colhead{IC (0.5--1)}\hspace{-2cm} & \colhead{} &	\colhead{SC (0--0.5)}\hspace{-2cm} & \colhead{} \\
\colhead{ObsId}	& \colhead{13285}	& \colhead{11790}		& \colhead{13285}		& \colhead{11790}	& \colhead{13285}		& \colhead{11790}}
\startdata
Exposure (ks)	& 9.84	& $8.27$\tablenotemark{*}	& 5.38		& $4.43$\tablenotemark{*}	& 4.47		& $3.79$\tablenotemark{*}	\\[1.5ex]
Total Counts	& 54	& 58		& 35		& 29		& 19	  	& 29		\\[1.5ex]
Count rate (cts/ks) & 5.5 $\pm$ 0.8& 7.0 $\pm$ 0.9	& $6.5 \pm 1.1$ & $6.5 \pm 1.2$	& 4.2 $\pm$ 1.0 & $7.7 \pm 1.4$	\\[1.5ex]
\enddata
\tablenotetext{*}{Effective exposure corrected for loss of counts in the chip gap for ObsId 11790 (total exposure was 19.86 ks).}
\end{deluxetable*}

\begin{deluxetable*}{@{}lccccccccc@{}}
\tabletypesize{\footnotesize}
\tablecolumns{7}
\tablewidth{0pt}
\tablecaption{Spectral parameters with 90\% confidence uncertainties for PSR J1311--3430.\label{specfitsJ1311}}
\tablehead{
\colhead{Phase range}	& \colhead{$N_{\rm H}$}	& \colhead{$\Gamma$}	&  \colhead{PL norm ($N_{-5}$)\tablenotemark{a}}	&   \colhead{$F^{\rm abs}_{0.3-7\,{\rm keV}}$\tablenotemark{b}}	&   \colhead{$F^{\rm unabs}_{0.3-7\,{\rm keV}}$\tablenotemark{b}}	& \colhead{Statistics}	\\
\colhead{}	& \colhead{($10^{20}$ cm$^{-2}$)}	& \colhead{}	&  \colhead{} & \colhead{} &  \colhead{}	& \colhead{Cstat/dof;\,$\chi^2_\nu$}	}
\startdata
Full orbit	& $5^{\rm F}$	& $1.39_{-0.26}^{+0.27}$	& $1.18_{-0.25}^{+0.30}$	& $8.1_{-1.3}^{+1.6}$	& $8.6_{-1.3}^{+1.5}$	& $82/101; 0.75$	\\
\midrule
IC phases	& $5^{\rm F}$	& $1.04_{-0.36}^{+0.36}$	& $0.99_{-0.30}^{+0.38}$	& $9.8_{-2.2}^{+2.7}$ 	& $10.2_{-2.2}^{+2.7}$	& $59/59; 0.99$	\\
\midrule
SC phases	& $5^{\rm F}$	& $1.96_{-0.43}^{+0.45}$	& $1.50_{-0.42}^{+0.51}$	& $6.7_{-1.5}^{+1.8}$	& $7.7_{-1.7}^{+2.0}$	& $33/44; 0.61$	\\
\enddata
\tablecomments{Spectra from ObsID 11790 and 13285 are simultaneously fit with only their parameters tied-up.}
\tablenotetext{a}{PL normalization in units of $10^{-5}$  photons cm$^{-2}$ s$^{-1}$ keV$^{-1}$ at 1 keV.}
\tablenotetext{b}{$F^{\rm abs}_{0.3-7\,{\rm keV}}$ and $F^{\rm unabs}_{0.3-7\,{\rm keV}}$ are absorbed and unabsorbed fluxes, respectively, in units of $10^{-14}$ erg cm$^{-2}$ s$^{-1}$.}
\tablenotetext{F}{Parameter fixed.}
\end{deluxetable*}

\begin{figure*}[ht]
\captionsetup[subfigure]{labelformat=empty}
\subfloat[]{\label{fig:J1311Sa}\includegraphics[width = 60mm]{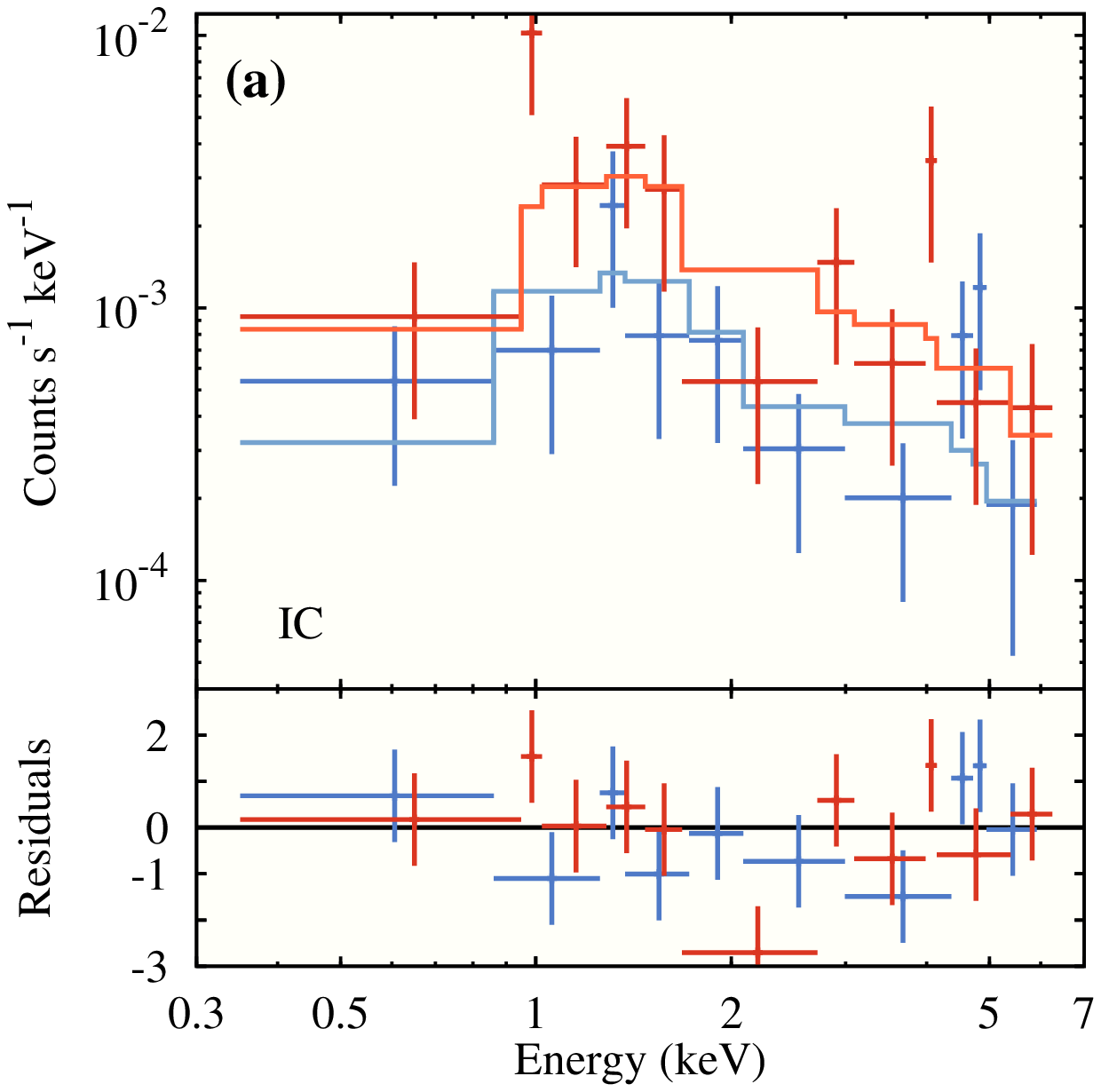}}
\subfloat[]{\label{fig:J1311Sb}\includegraphics[width = 60mm]{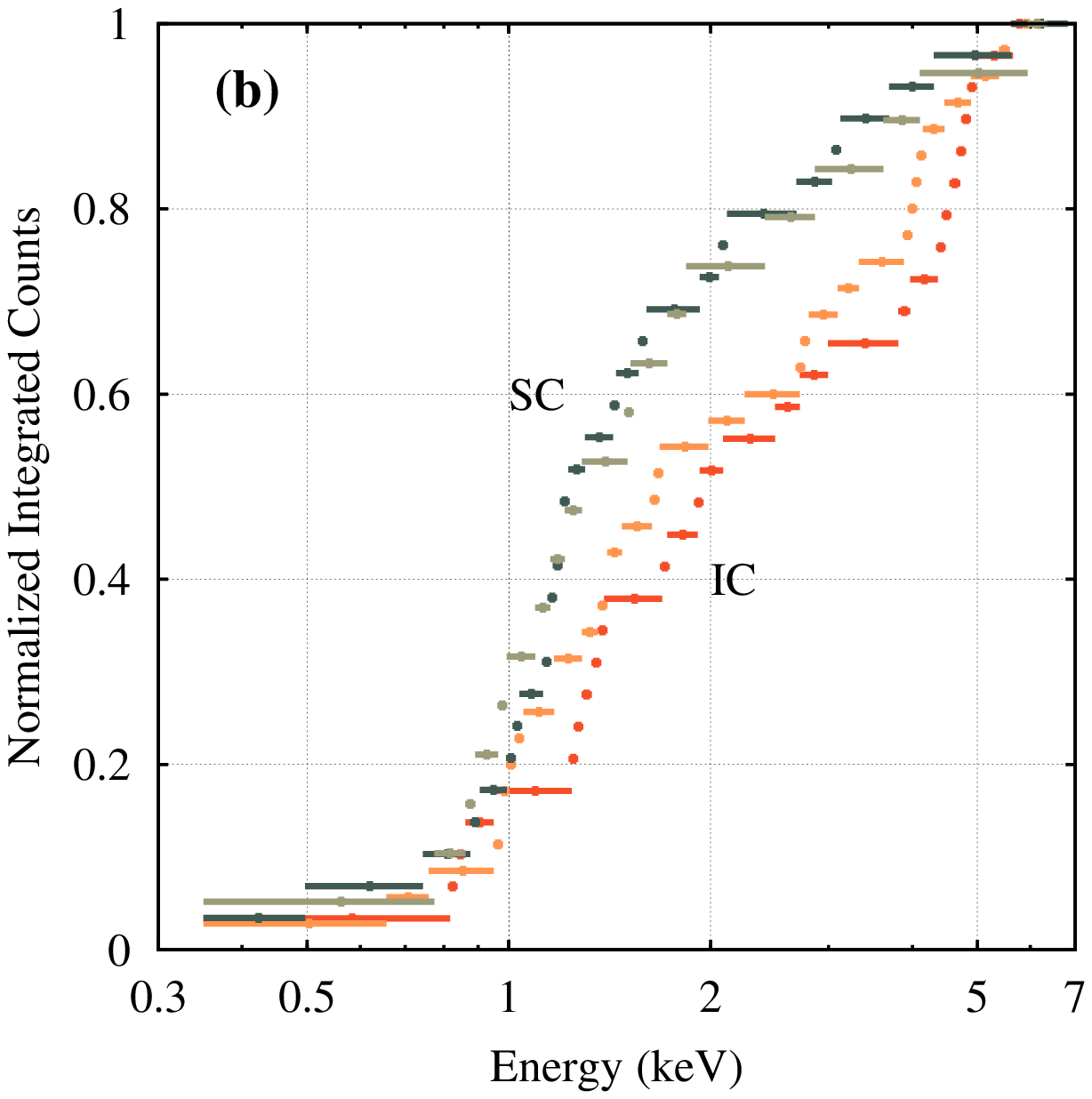}}
\subfloat[]{\label{fig:J1311Sc}\includegraphics[width = 60mm]{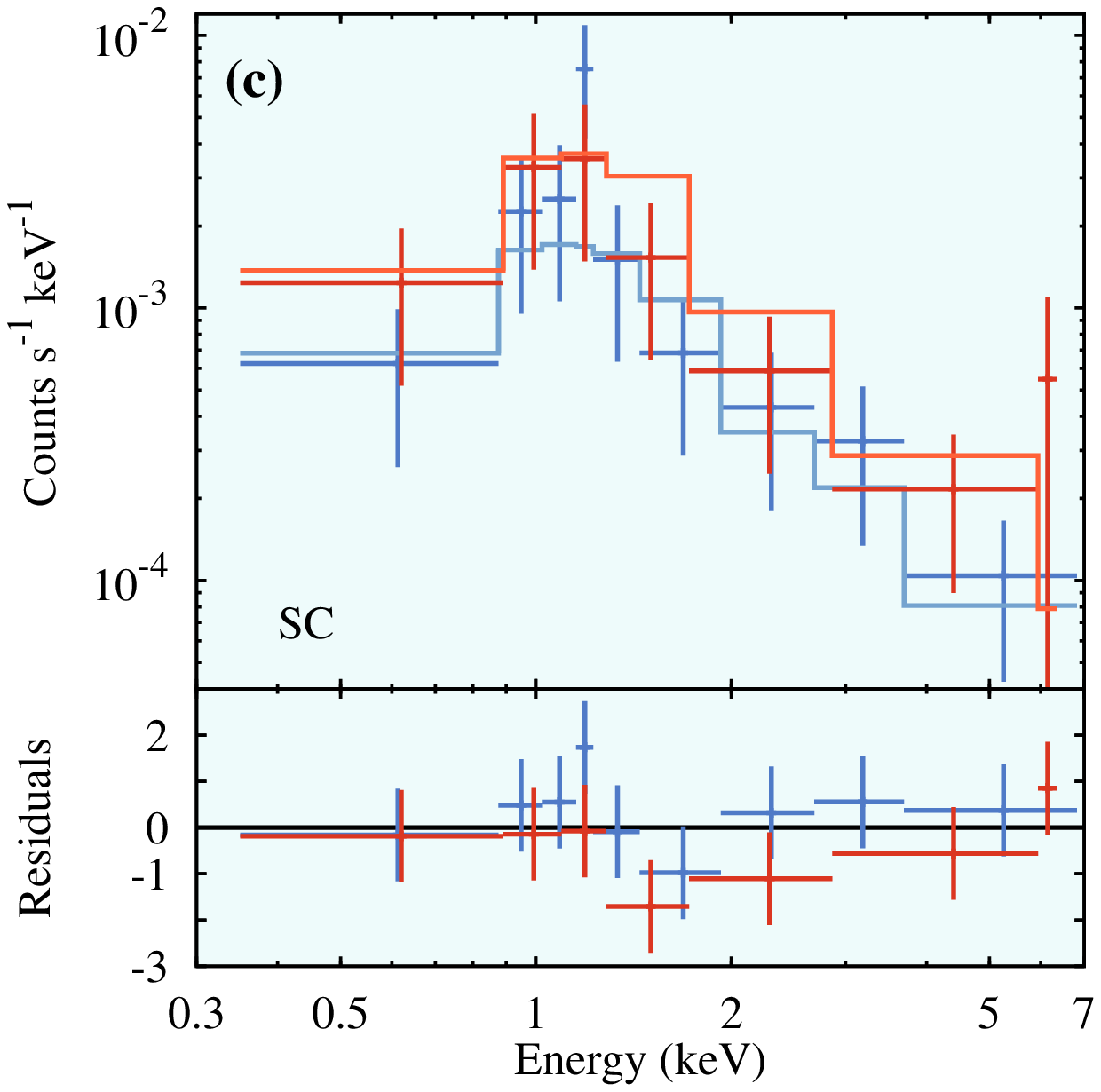}} \\
\subfloat[]{\label{fig:J1311Sd}\includegraphics[width = 60mm]{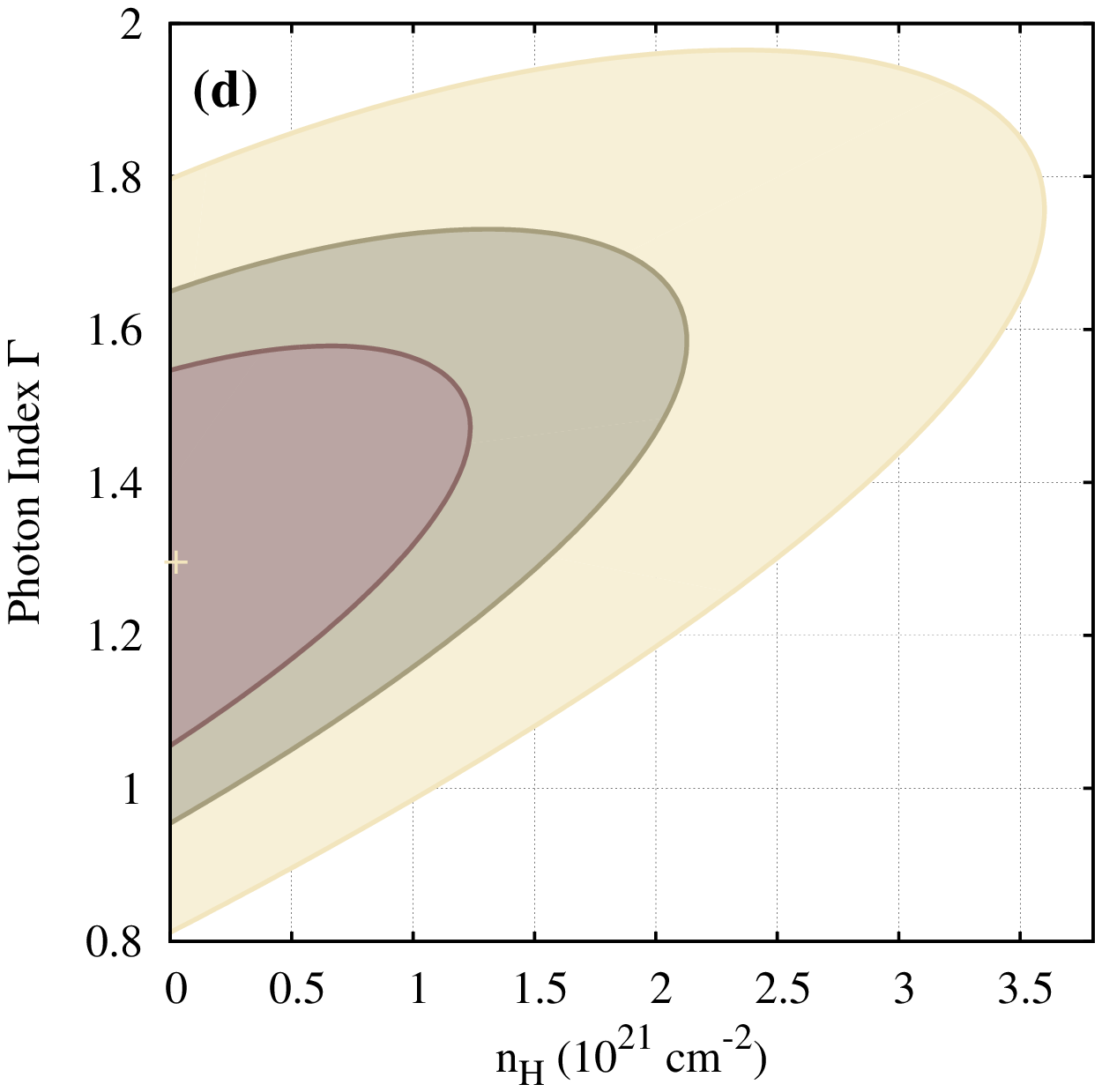}}
\subfloat[]{\label{fig:J1311Se}\includegraphics[width = 60mm]{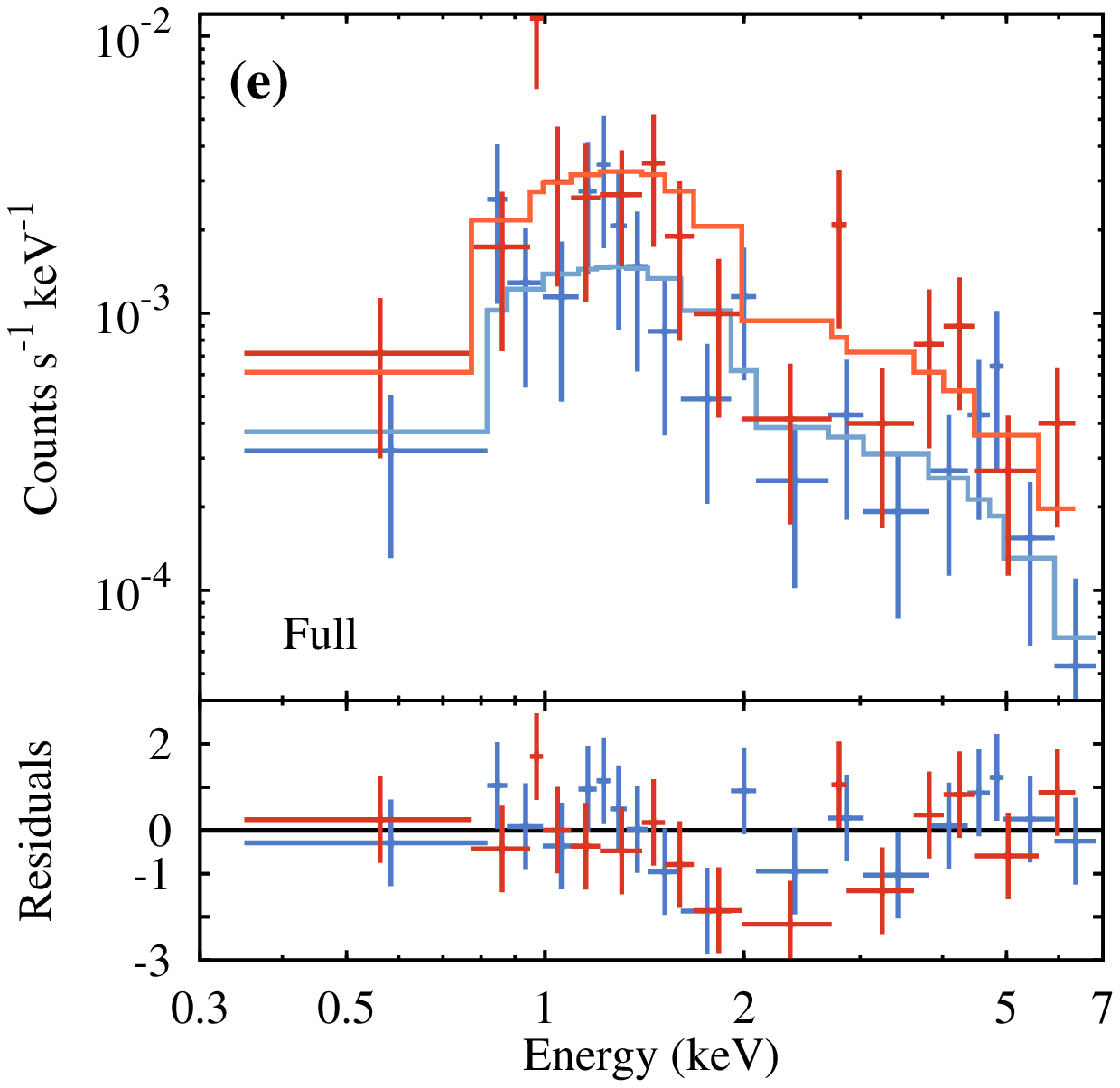}}
\subfloat[]{\label{fig:J1311Sf}\includegraphics[width = 60mm]{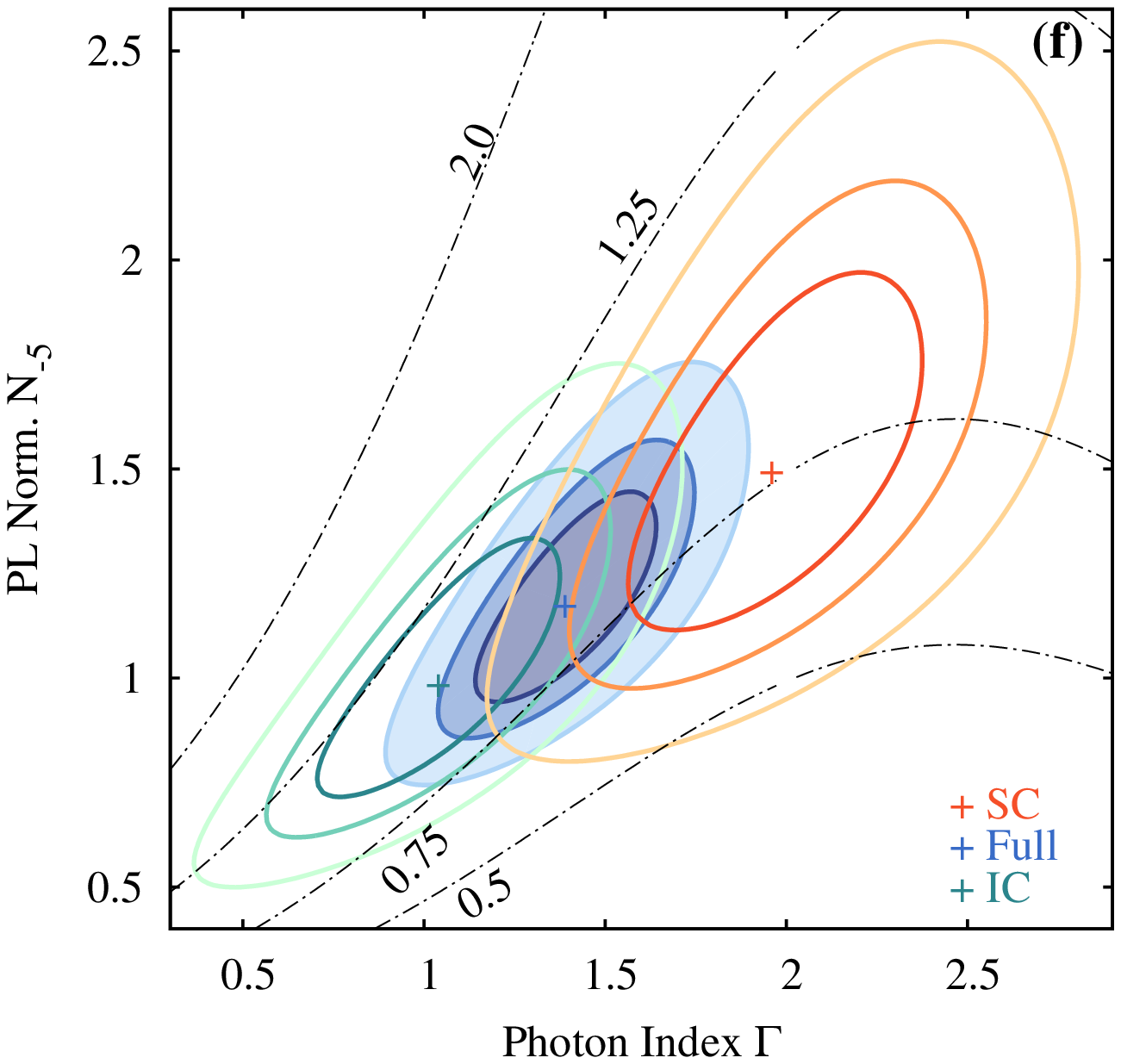}} \\
\caption{Spectral analysis of J1311--3430. Panel \protect\subref{fig:J1311Sb} shows the normalized cumulative counts for the IC (red) and SC (green) phase-resolved spectra (dark: MJD 55276, light: MJD 56013), while panels \protect\subref{fig:J1311Sa} and \protect\subref{fig:J1311Sc} show PL fits to IC and SC spectra, respectively (red: MJD 55276, blue: MJD 56013).
Panel \protect\subref{fig:J1311Se} shows PL fits to the phase-integrated spectra.
Panel \protect\subref{fig:J1311Sd} shows 68\%, 95\% and 99\% confidence contours in the $\Gamma - N_{\rm H}$ plane for PL fit to the phase-integrated spectra, whereas panels \protect\subref{fig:J1311Sf}  show $N_{-5} - \Gamma$ confidence contours for IC (green), full (blue) and SC (red) spectral fits.
Lines of constant flux are overplotted on $N_{-5} - \Gamma$ contours with values quoted in units of $10^{-13}$ erg cm$^{-2}$ s$^{-1}$.}
\label{J1311spectrals}
\end{figure*}

\clearpage

\section{Other BWP in the X-rays}
Over a dozen BWPs and BWP candidates have been targeted using {\em Chandra} and \xmm (Table \ref{bwps-xpar}).
However, very few systems have been observed with the necessary exposure depth for a detailed timing/spectral analysis.
We adopt a simple, uniform spectral extraction and fitting procedure to analyze the {\em Chandra} data on 12 BWPs (and candidates), and compare their phase-integrated emission characteristics.
The data are reprocessed using the standard CIAO tasks.
We restrict the event extraction to the 0.3 -- 7 keV energy range and limit the aperture radius to $2\farcs5$.

Each spectrum was first fitted with a simple PL model.
Spectra with soft ($\Gamma \gtrsim 3$) PL fits were also fit with a BB model.
The spectrum of J1959+2048 (= B1957+20) has enough counts to allow a combined PL+BB fit.
Ten out of the 12 spectra have $\lesssim 300$ counts, so we use their unbinned spectra and fit models using C-statistic, and the other 2 are binned ($\geq 15$ counts per bin) and fit using $\chi^2$ statistic.
The absorption column density $N_{\rm H}$ is initially set as a free parameter but subsequently fixed at the best fit value to avoid over-fitting of the low quantity spectra.

Most of the pulsar's X-ray data have been analyzed in previous works (see references in Table \ref{bwps-xpar}), and the spectral fits are consistent with our results. 
For pulsars J2241--5236 and J2214+3000, only BB spectral fits were available so, we provide statistically acceptable PL fits. 
For pulsars J1731--1847 and J2047+1053, no previous analyses have been presented, and for J1311--3430 we introduce data from newer observations.

\section{Discussion and Conclusions}

\subsection{Distributions of BWP properties}

\begin{figure*}[t]
\includegraphics[width = 180mm]{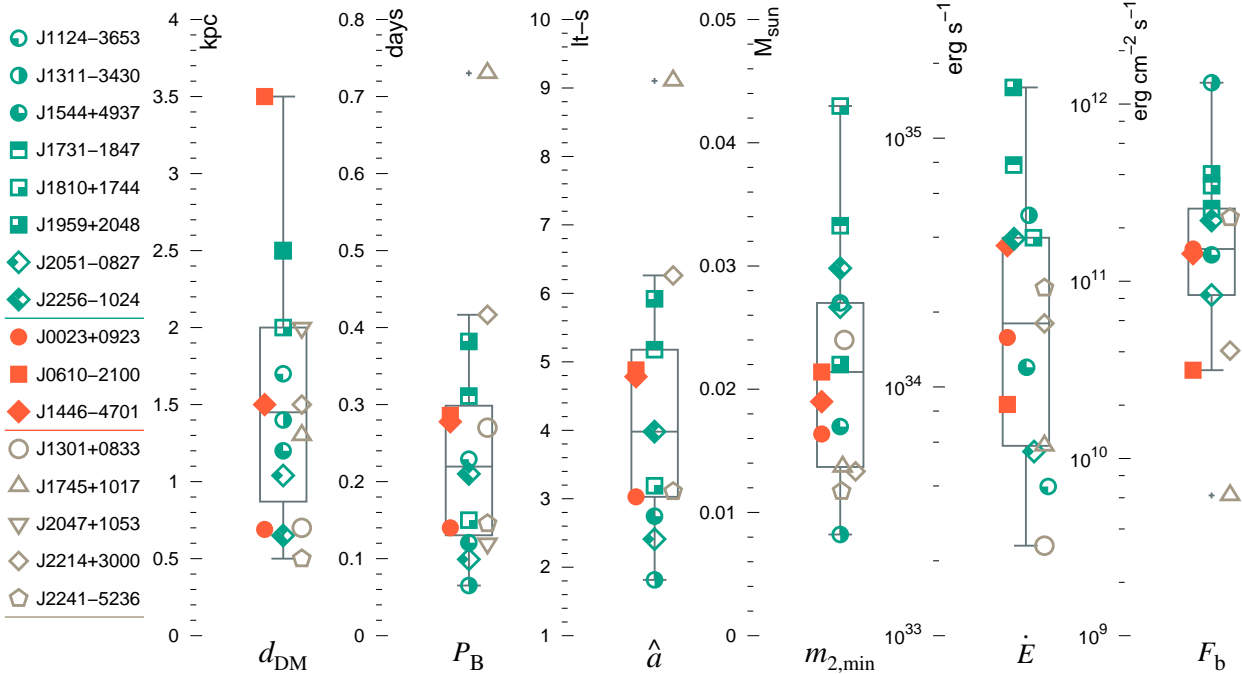}
\caption{Multi-dimensional charting of various BWP properties (references in Table \ref{bwps-xpar}): DM distance $d_{\rm DM}$), binary period $P_{\rm B}$, empirical binary separation $\hat{a}$, minimum companion mass ($m_{\rm 2,min} = m_2\,\sin i$), spin-down power ($\dot{E}$), and intra-binary flux ($F_{\rm b}$).
Box plots represent the distributions, with middle bars at the median values, boundaries at the quartiles, and `whiskers' that extend up to the last data point that falls within $1.5\times$ the inter-quartile range (the height of the box).
The 8 known eclipsing BWPs are represented in green with partially filled markers, the 3 likely non-eclipsers are in red fully-filled markers, and the rest of the BWPs with unknown status are in unfilled grey markers.
Pulsars with known proper motion and Shklovskii effect corrected $\dot{E}$ values are J0610-2100, J1311-3430, J1446-4701, J1745+1017, J1959+2048 and J2051-0827.
}
\label{distributions}
\end{figure*}

There are 16 known (plus candidate BWP J1653.6--0159) BWPs identified either through detection of eclipses in the radio, or through low companion masses and short orbital periods.
The small sample does not lend itself to robust population analysis but nevertheless shows some patterns worth exploring.
The distribution of various BWP system parameters (Figure \ref{distributions}) shows that the sample of BWPs has the following median characteristics:
\begin{itemize}[leftmargin=*,label={}]
\itemsep0.01em 
  \item orbital period $P_{\rm B} = 0.22$ days,
  \item empirical binary separation $\hat{a} = 4$ lt-s,
  \item minimum companion mass $m_2\,\sin i = 0.021\,M_\odot$, and
  \item pulsar spin-down power $\dot{E} = 1.8 \times 10^{34}$ erg s$^{-1}$. 
\end{itemize}
As an estimate for the intra-binary distance, $a = [1+(m_1/m_2)]\,a_1$, we use $\hat{a} = [1+m_1/(m_2\, \sin\,i)] \times (a_1 \sin\,i)$.
The neutron star mass is assumed to be $1.4\,M_\odot$, although there is some emerging evidence for higher mass pulsars ($m_1 \sim 2\,M_\odot$) in such systems (\citealt{2011ApJ...728...95V}; \citealt{2012ApJ...760L..36R}).
Even though $\hat{a}$ is technically the minimum binary separation, for BWPs, it varies from the actual intra-binary distance by $<$ 10\%, for $i \gtrsim 10^\circ$ and $m_2 \lesssim 0.15 M_\odot$.
Unless a significant fraction of the sample BWPs have $i < 10^\circ$, the distribution of $a$ should be offset only marginally towards higher values compared to the distribution of $\hat{a}$.

The minimum companion masses are found assuming $i = 90^\circ$.
For eclipsing systems, if we assume an inclination of $> 60^\circ$, the true companion masses will be a factor of $\lesssim 1.2$ higher, while, for non-eclipsers, an inclination of $\sim 30^\circ$ will imply a factor of 2 higher true companion masses.
The distribution of true companion masses could be significantly different than that of minimum companion masses, with the maximum value possibly over 0.05 M$_\sun$.

For each system, one can roughly estimate the spin-down flux available at its companion's location, $F_b = \dot{E}/[4\,\pi\,\hat{a}^2]$, which we call the intra-binary flux.
The intra-binary flux values in this sample of BWPs span two orders of magnitude, and they could serve as a simple measure of the pulsar wind/radiative power that produces the intra-binary shocks and causes companion surface irradiation/ablation. 
It is important to note that, with the exception of pulsars J0610-2100, J1311-3430, J1446-4701, J1745+1017, J1959+2048 and J2051-0827, the $\dot{P}$ values were not corrected for the Shklovskii effect \citep{1970SvA....13..562S} because of unknown proper motion.
The resulting bias in the measured $\dot{P}$ values leads to overestimation of the pulsars' $\dot{E}$ values.

\subsection{BWP J1446--4701}

Detected at 1.4 GHz without orbital eclipses, the pulsar lacks the necessary coverage at lower radio frequencies to definitely classify it as a non-eclipsing binary.
If it fails to show eclipses or DM variations, it could either be due to weakness of its intra-binary shock or a low orbital inclination.
The hints of X-ray orbital modulations found in the \xmm data lack the statistical significance to constrain the inclination angle.
For low orbital inclinations, the emission from the pulsar and the intra-binary shock should be simultaneously observable throughout the orbit, and hence the total emission should be unmodulated.

BWP J1446--4701 has an X-ray luminosity $L_{\rm 0.1 - 10 keV}^{\rm unabs} = 2.4_{-1.2}^{+3.8} \times 10^{31}$ erg s$^{-1}$ (90\% uncertainty limits), which is higher than the median $L_{\rm X}$ value of the BWP sample.
The phase-integrated and phase-resolved spectra allow soft, $\Gamma \approx 3$, PL fits with a hint of spectral hardening near superior conjunction.
Such soft PL fits have been observed in other MSPs but those MSPs have at least an order of magnitude lower X-ray luminosity than J1446--4701 (see Table 2 in \citealp{2007ApJ...664.1072P}).
The alternative PL+BB fit with $\Gamma = 1.5$ is motivated by the simple BWP emission model which combines emission from pulsar polar cap and the intra-binary shock.
For a sufficiently large inclination, Doppler boosted X-ray flux excesses are expected at SC phases.
SC phases in J1446--4701 do show as increased PL component compared to the IC phases but only at the 99.4\% confidence level.
A higher S/N data would allow differentiation of the PL from BB components in the spectra at different orbital phases, and help detect flux modulations due to the intra-binary shock.

The non-detection of the optical companion to J1446--4701 sets an absolute B magnitude limit $M_{\rm B} > 17.1$.
There are three other pulsars having similar intra-binary fluxes, with detections in the optical, the eclipsers J1544+4937 with $M_{\rm B} = 17.7$ \citep{2014ApJ...791L...5T} and J2256--1024 with $M_{\rm B} = 19.2$ \citep{2013ApJ...769..108B}, and the non-eclipser J0023+0923 with $M_{\rm B} = 18.1$ \citep{2013ApJ...769..108B}.
If PSR J1446--4701 is in-fact a non-eclipser, similar to J0023+0923, it may have to be targeted down to a sensitivity limit of $m_{\rm B} \gtrsim 25$ to detect its irradiated side.
Currently, the lack of radio eclipses, a low-confidence X-ray flux/spectral modulation, a soft X-ray spectrum, and non-detection of the companion in the optical point towards a low orbital inclination for BWP J1446--4701.

\subsection{BWP J1311--3430}

BWP J1311--3430 has been observed in the radio, optical, X-rays and $\gamma$-rays.
The system is extreme in its binary parameters, having the lowest binary separation and minimum companion mass of all known BWPs (candidate BWP J1653.6-0159 could beat the record if confirmed). 
These extreme properties have allowed detections of $\gamma$-ray orbital modulations and X-ray flares, not yet detected in other BWP systems.
X-ray flares with a factor of $\sim 10$ flux variation, at timescales of 10 ks, have been detected in {\em Suzaku} observations in August 2009 and August 2011 (\citealp{2011ApJ...729..103M}; \citealp{2012ApJ...757..176K}).

The pulsar's hard, $\Gamma \approx 1.4$, PL spectral fit is consistent with the results in previous {\em Suzaku} (\citealp{2011ApJ...729..103M} \citealp{2012ApJ...757..176K}) and {\em Chandra} \citep{2012ApJ...756...33C} data analyses.
The emission is more likely the result of an intra-binary shock rather than the typically softer pulsar magnetosheric emission of regular non-interacting MSPs.
Our flux estimates, although consistent between the two {\em Chandra} observations from March 2010 and March 2012, are significantly lower than the estimates from the {\em Suzaku} observations.
Our flux estimates from {\em Chandra} data, $F^{\rm unabs}_{0.5-10\,{\rm keV}} = 6.3^{+1.4}_{-0.9} \times 10^{-14}$ erg cm$^{-2}$ s$^{-1}$ (quoted $1 \sigma$ uncertainties) is $\approx 7.7 \sigma$ lower than corresponding estimates from {\em Suzaku} observations, $F^{\rm unabs}_{0.5-10\,{\rm keV}} = 2.77^{+0.31}_{-0.24} \times 10^{-13}$ erg cm$^{-2}$ s$^{-1}$ \citep{2012ApJ...757..176K}.

Barring any cross-calibration discrepancy between the two instruments, the variability is most-likely intrinsic to the system.
The intrinsic X-ray variability between the observations could either be due to inhomogeneities in the ablated material and the Rayeligh-Taylor instabilities at the shock front, or due to activity in the stellar corona.
If the flaring activity is localized to the SC phases with higher relative confidence, it may indicate that the flares originate at the shock.
Better quality data in both flaring and quiescent states are needed to establish the cause of the X-ray variability.

Due to the system's extreme orbital parameters and the strong orbital modulations in the radio and optical, it is reasonable to expect X-ray orbital modulations.
Despite gaps in the observation, there is some evidence of X-ray orbital flux modulation in the {\em Suzaku} observations, but the phase of count-rate excess is different in different observations \citep{2012ApJ...757..176K}.
The latter {\em Chandra} observation samples the X-ray photons more uniformly than the previous one, but the short observation does not allow a high S/N timing analysis.
Accordingly, only a hint of X-ray depletion at SC phases is seen in our data.

\subsection{The BWP sample in the X-rays}

\begin{figure}[H]
\includegraphics[width = 85mm]{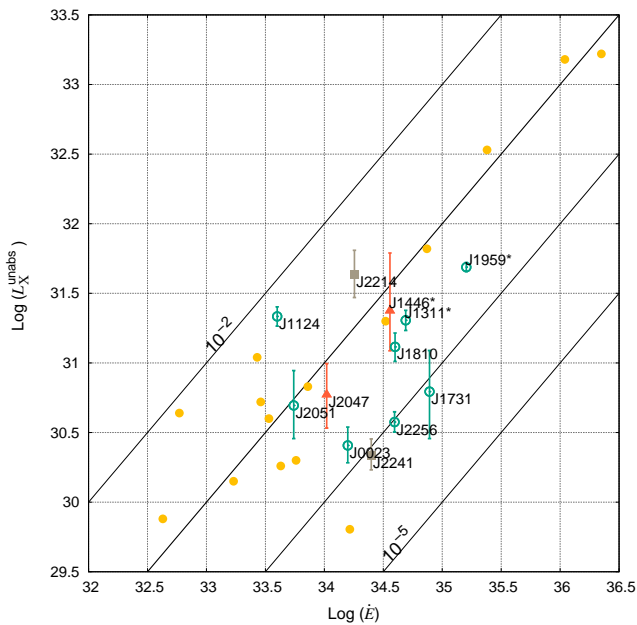}
\caption{Spin-down power versus 0.1 -- 10 keV luminosity for 12 BWPs analyzed in this paper and 15 non-BWP MSPs from \cite{2007ApJ...664.1072P}.
Labels of pulsars with known proper motion and Shklovskii effect corrected $\dot{E}$ values suffixed with asterisk (*).
The uncertainties reflect only the flux uncertainties and ignore the distance uncertainties.
The sample MSPs are represented with yellow dots while BWPs are classified into eclipsers (green circles), non-eclipsers (red triangles) and ones with unknown status (grey squares).
Diagonals represent lines of constant X-ray efficiency, $\eta_{\rm X} = L_{\rm X}/\dot{E}$, between $10^{-5} - 10^{-2}$}
\label{BWPXrays}
\end{figure}

The BWPs detected in the X-rays span a wide range of system parameters as seen in Figure \ref{distributions}.
It is fairly reasonable to assume that the X-ray emission, which we expect to have partial contribution from the intra-binary shock, must depend on the orbital parameters and the intrinsic pulsar properties.
Comparison of the BWP X-ray and spin-down luminosities with those of other MSPs shows that the BWPs occupy a narrow region around higher median $\dot{E}$ and $L_{\rm X}$ (Figure \ref{BWPXrays}).
There also seems to be an absence of correlation between the the BWP $\dot{E}$ and $L_{\rm X}$.
If companion ablation in recycled-MSP binaries is a phenomenon that occurs during a short period in MSP evolution, one can expect a similarly narrow distribution of $\dot{E}$.
Given the narrow distribution of $\dot{E}$, the uncertainties in the X-ray flux and distance estimates need to be much lower to establish correlation or lack thereof.
We should note, however, that majority of our sample of BWPs lack proper motion measurements, hence the Shkolvskii corrections to their $\dot{E}$ values are unknown.
This may change the $\dot{E}$ and $L_{\rm X}$ distributions significantly.

There seems to be clustering of pulsars with softer ($\Gamma >  2.5$) spectra at relatively lower luminosities ($L_{\rm 0.1 - 10\,keV} < 10^{31}$ erg s$^{-1}$).
The pattern, if real, could be the result of lower intra-binary flux from systems exhibiting soft X-ray spectra.
Interestingly, these pulsars with low X-ray luminosities are also clustered together at the lowest $\gamma$-ray luminosities in the BWP sample.
PSR J1731--1847, with $\approx 6$ counts in $\approx 10$ ks of ACIS-I exposure, is poorly constrained ($\Gamma = 1.9\pm 1.4$) and could turn out to have a softer spectrum in line with the clustering seen above.
J1446--4701 is the only outlier, which, despite its soft ($\Gamma \approx 2.9$) PL fit, falls among the higher luminosity group of BWPs.

The X-ray luminosities or spectral slopes of BWPs do not show any correlation with the system's binary separation or orbital period in our small sample.
It is unclear whether such a lack of correlations between the pulsar properties and their X-ray emission is due to a small sample size and, as true for most systems, low data quality.
Higher quality data and inclusion of more systems can help understand or improve these results.
If such anomalous results persist, we will have to revise our models of BWP interactions and their emission mechanisms.

\begin{deluxetable*}{@{}lccccccccc@{}}
\tabletypesize{\footnotesize}
\tablecolumns{10}
\tablewidth{0pt}
\tablecaption{X-ray observational parameters for BWPs detected by {\sl Chandra}.\label{bwps-xpar}}
\tablehead{
\colhead{Pulsar}       & \colhead{ObsId}        & \colhead{$P_{\rm B}$} & \colhead{Exposure} & \colhead{Orbital} & \colhead{Counts} & \colhead{Count Rate}     & \colhead{Distance} & \colhead{log($\dot{E}$)}         & \colhead{Refs.$^b$} \\
\colhead{}             &  \colhead{}            & \colhead{(hr)}        & \colhead{(ks)}     & \colhead{Coverage\tablenotemark{a}}  &   \colhead{}     & \colhead{(cts ks$^{-1}$)}  & \colhead{(kpc)}      & \colhead{(erg s$^{-1}$)} &  \colhead{} }
\startdata
J1124--3653   & 13722        & 5.50        & 19.95    & 1.01             & 135    & $6.76\pm0.58$  & 1.7      & 33.60        & 1, 2     \\[1.1EX]
J1311--3430\tablenotemark{*}   & 11790,13285        & 1.56        & 18.11     & 2.43             & 112     & $7.01\pm0.92$  & 1.4      & 34.69        & 3, 4, 5    \\[1.1EX]
J1731--1847   & 13290        & 7.47        & 9.457    & 0.35             & 6      & $0.60\pm0.25$  & 2.5     & 34.89        & 5, 6  \\[1.1EX]
J1810+1744   & 12465        & 3.6        & 20.45    & 1.58             & 52     & $2.53\pm0.35$  & 2.0     & 34.60        &  1, 2   \\[1.1EX]
J1959+2048\tablenotemark{*}   & 01911,09088  & 9.17        & 209.5    & 6.35             & 1521   & $7.25\pm0.19$  & 2.5     & 35.20        & 7, 8     \\[1.1EX]
J2051--0827\tablenotemark{*}   & 10106--10110 & 2.38        & 44.29    & 5.17             & 42     & $0.93\pm0.15$  & 1.0     & 33.74        & 9, 10     \\[1.1EX]
J2256--1024   & 12467        & 5.04        & 19.8     & 1.09             & 135    & $6.81\pm0.59$  & 0.7     & 34.60        & 1, 2    \\[2.1EX]
J0023+0923   & 12464,14363  & 3.36        & 16.7      & 1.38            & 57     & $3.40\pm0.45$  & 0.7     & 34.20        & 1, 2     \\[2.1EX]
J2047+1053   & 14472        & 2.88        & 17.74    & 1.71             & 17     & $0.95\pm0.23$  & 2.0     & \ldots        & 5, 11  \\[1.1EX]
J2214+3000   & 11788        & 10.00        & 18.139   & 0.50             & 79     & $4.01\pm0.45$  & 1.5     & 34.26        & 12     \\[1.1EX]
J2241--5236   & 11789        & 3.50        & 19.246   & 1.53             & 81     & $4.06\pm0.45$  & 0.5     & 34.40        & 13     \\[2.1EX]
J1653.6--0159 & 11787        & 1.25        & 20.8     & 4.63             & 346    & $16.63\pm0.90$ & \ldots       & \ldots        & 3, 14   \\[1.1EX]
\enddata
\tablecomments{The first 7 pulsars are known eclipsers, J0023+0923 is most-likely a non-eclipser, and the rest have unknown status.}
\tablenotetext{a}{Number/fraction of BWP orbits covered in the observations.}
\tablenotetext{*}{Pulsars with known proper motion and Shklovskii effect corrected $\dot{E}$ values.}
\tablenotetext{b}{References: 
(1) \citealp{2011AIPC.1357...40H}, (2) \citealp{2014ApJ...783...69G},
(3) \citealp{2012ApJ...756...33C}, (4) \citealp{2013ApJ...763L..13R}, (5) this work,
(6) \citealp{2011MNRAS.416.2455B},
(7) \citealp{1988Natur.333..237F}, (8) \citealp{2012ApJ...760...92H},
(9) \citealp{1996ApJ...465L.119S}, (10) \citealp{2013MNRAS.430..571E},
(11) \citealp{2012arXiv1205.3089R},
(12) \citealp{2011ApJ...727L..16R},
(13) \citealp{2011MNRAS.414.1292K},
(14) \citealp{2014ApJ...793L..20R}.}
\end{deluxetable*}

\begin{deluxetable*}{@{}lcccccc@{}}
\tabletypesize{\footnotesize}
\tablecolumns{10}
\tablewidth{0pt}
\tablecaption{Spectral fits to BWPs' X-ray spectra.\label{bwps-xfit}}
\tablehead{
\colhead{Pulsar} & \colhead{Model} & \colhead{$N_{\rm H}$} & \colhead{$\Gamma/kT^{\rm a}$} & \colhead{$\log (F_{\rm abs})$\tablenotemark{a}} & \colhead{$\log (F_{\rm unabs})$\tablenotemark{a}} & \colhead{Cstat/dof ($\chi_\nu^2$)} \\
\colhead{} & \colhead{} & \colhead{($10^{20} {\rm cm}^{-2}$)} & \colhead{(\ldots/keV)} & \colhead{(erg cm$^{-2}$ s$^{-1}$)} & \colhead{(erg cm$^{-2}$ s$^{-1}$)} & \colhead{} }
\startdata
Eclipsers &&&&&&\\[0.1EX]
\cmidrule(r){1-1}
J1124--3653   & PL    & $< 9.1$                & $1.6_{-0.2}^{+0.2}$                      & \multicolumn{2}{c}{$-13.20_{-0.07}^{+0.07}$} & 83.6/98 (0.87)  \\[1.1EX]
J1311--3430   & PL    & $< 14$			& $1.3_{-0.3}^{+0.3}$			& \multicolumn{2}{c}{$-13.07_{-0.08}^{+0.07}$} & 81.5/100 (0.74)  \\[1.1EX]
J1731--1847   & PL    & $< 136$                 & $1.9_{-1.3}^{+1.5}$                       & \multicolumn{2}{c}{$-14.08_{-0.34}^{+0.30}$} & 1.3/6 (0.32)    \\[1.1EX]
J1810+1744   & PL    & $17_{-17}^{+27}$  & $2.2_{-0.4}^{+0.4}$                      & $-13.74_{-0.12}^{+0.12}$ & $-13.56_{-0.10}^{+0.10}$ & 27.6/43 (0.70)  \\[1.1EX]
J1959+2048   & PL    & $16.5_{-4.0}^{+4.4}$    & $2.0_{-0.1}^{+0.2}$                      & $-13.32_{-0.03}^{+0.03}$ & $-13.19_{-0.03}^{+0.03}$ & \ldots/73 (1.12)      \\[1.1EX]
             & PL+BB & $5.4_{-5.4}^{+9.6}$  & $1.2_{-0.3}^{+0.2}/0.29_{-0.03}^{+0.03}$ & $-13.29_{-0.03}^{+0.03}$ & $-13.26_{-0.03}^{+0.03}$ & \ldots/72 (1.03)       \\[1.1EX]
J2051--0827   & PL    & $43_{-32}^{+45}$   & $4.1_{-0.7}^{+0.7}$                      & $-14.37_{-0.12}^{+0.11}$ & $-13.42_{-0.24}^{+0.25}$ & 28.4/34 (0.69)  \\[1.1EX]
J2256--1024   & PL    & $10_{-10}^{+13}$  & $2.9_{-0.3}^{+0.3}$                  & $-13.35_{-0.06}^{+0.06}$ & $-13.13_{-0.07}^{+0.07}$ & 55.6/82 (0.92)  \\[2.1EX]
\cmidrule(r){1-1}
Non-eclipsers &&&&&&\\[0.1EX]
\cmidrule(r){1-1}
J0023+0923   & PL    & $8.7_{-8.7}^{+2.4}$  & $3.3_{-0.5}^{+0.5}$                      & $-13.61_{-0.10}^{+0.09}$ & $-13.35_{-0.12}^{+0.13}$ & 23.1/46 (0.57)  \\[1.1EX]
             & BB    & $< 4.8$                 & $0.22_{-0.03}^{+0.04}$                      & \multicolumn{2}{c}{$-13.73_{-0.10}^{+0.10}$} & 34.4/46 (0.88)  \\[2.1EX]
J1446--4701	& PL	& $17^{+10}_{-8}$	& $2.9^{+0.5}_{-0.4}$	& $-13.83_{-0.07}^{+0.07}$	& $-13.05_{-0.29}^{+0.42}$	&	\ldots/56 (1.13) \\[1.1EX]
\cmidrule(r){1-1}
Unknown Status &&&&&&\\[0.1EX]
\cmidrule(r){1-1}
J2047+1053   & PL    & $< 58$                & $0.87_{-0.68}^{+0.68}$                      & \multicolumn{2}{c}{$-13.91_{-0.24}^{+0.22}$} & 16.9/15 (1.21)  \\[1.1EX]
             & BB    & $< 25$             & $0.27_{-0.05}^{+0.06}$                      & \multicolumn{2}{c}{$-14.39_{-0.12}^{+0.11}$} & 27.6/34 (0.64)  \\[1.1EX]
J2214+3000   & PL    & $19_{-19}^{+28}$   & $3.8_{-0.5}^{+0.5}$                      & $-13.36_{-0.09}^{+0.09}$ & $-12.80_{-0.16}^{+0.18}$ & 47.8/ 54 (0.83) \\[1.1EX]
J2241--5236   & PL    & $11_{-11}^{+23}$  & $2.8_{-0.4}^{+0.4}$                      & $-13.36_{-0.08}^{+0.08}$ & $-13.14_{-0.11}^{+0.12}$ & 48.3/59 (1.38)  \\[1.1EX]
             & BB    & $< 7.0$                & $0.24_{-0.03}^{+0.04}$                      & \multicolumn{2}{c}{$-13.46_{-0.09}^{+0.09}$} & 58.5/54 (0.99)  \\[1.1EX]
             &       &                         &                                             &                          &                          &                 \\
\cmidrule(r){1-1}
Candidate BWP &&&&&&\\[0.1EX]
\cmidrule(r){1-1}
J1653.6--0159 & PL    & $21_{-15}^{+19}$  & $1.7_{-0.2}^{+0.2}$                      & $-12.71_{-0.05}^{+0.05}$ & $-12.60_{-0.04}^{+0.04}$ & \ldots/20 (0.66)      \\[1.1EX]
\enddata
\tablenotetext{a}{$F_{\rm abs}$ and $F_{\rm unabs}$ are the absorbed and unabsorbed flux in the 0.15 -- 10 keV range, respectively.}
\end{deluxetable*}

\clearpage
\acknowledgments

We thank Eric Feigelson for valuable discussion and suggestion on statistical techniques, and the {\em Chandra} helpdesk for important clarifications regarding the {\em Chandra} data analysis.
We also thank the referee for careful reading of the manuscript and very useful remarks.
Part of this work is based on observations obtained with {\sl XMM-Newton}, an ESA science mission with instruments and contributions directly funded by ESA Member States and NASA.
This work was supported by the Penn State ACIS Instrument Team Contract SV4-74018, issued by the Chandra X-ray Center, which is operated by the Smithsonian Astrophysical Observatory for and on behalf of NASA under contract NAS8-03060.
The Guaranteed Time Observations (GTO) included here were selected by the ACIS Instrument Principal Investigator, Gordon P. Garmire, of the Huntingdon Institute for X-ray Astronomy, LLC, which is under contract to the Smithsonian Astrophysical Observatory; Contract SV2-82024. 
This work was partly supported by NASA grant NNX13AF02G.

Facilities: \facility{{\sl XMM-Newton}}, \facility{CXO}.
\bibliographystyle{apj}
\bibliography{blackwidows}

\end{document}